\documentclass[a4paper,twocolumn,11pt,accepted=2026-09-05]{quantumarticle}
\pdfoutput=1
\usepackage[utf8]{inputenc}
\usepackage[english]{babel}
\usepackage[T1]{fontenc}
\usepackage{graphicx}
\usepackage{microtype}
\usepackage{needspace}
\usepackage{amsmath}
\usepackage{amssymb}
\usepackage{bm}
\usepackage{braket}
\usepackage{physics}
\usepackage{dsfont}
\usepackage{relsize}
\usepackage{comment}
\usepackage[numbers]{natbib}
\usepackage{hyperref}
\hypersetup{hidelinks}

\DeclareRobustCommand{\redlined}[1]{#1}
\begin{document}
\title{Measuring a Quantum Measure Exceeding Unity}
\author[1]{Sanchari Chakraborti}
\author[2,3,5]{Rafael D. Sorkin}
\author[1,4]{Urbasi Sinha}
\email{usinha@rri.res.in}
\affil[1]{Raman Research Institute, C.V.Raman Avenue, Sadashivanagar, Bengaluru-560080, Karnataka, India}
\affil[2]{Perimeter Institute for Theoretical Physics, Waterloo, ON N2L 2Y5, Canada}
\affil[3]{Department of Physics, Syracuse University, Syracuse, NY 13244-1130, U.S.A.}
\affil[4]{Department of Physics and Astronomy, University of Calgary, Alberta T2N 1N4, Canada}
\affil[5]{School of Theoretical Physics, Dublin Institute for Advanced Studies, 10 Burlington Road, Dublin 4, Ireland}
\maketitle

\begin{abstract}
The history based formalism known as Quantum Measure Theory (QMT)
generalizes the concept of probability-measure so as to incorporate
quantum interference.  The resulting \textit{quantum measure} $\mu$ is
defined for arbitrary events (sets of histories), not just for
observables at a fixed moment of time. Thanks to interference effects,
$\mu$
can exceed unity, exhibiting its non-classical nature in a
particularly striking manner.  Here, in an optical experiment, we
illustrate an ancilla based filtering scheme that gives operational meaning to the quantum measure. For a specific photonic
event $E$, we report a measured value of \redlined{$\mu(E)=1.172^{+0.013}_{-0.019}$}, which within
errors agrees with the theoretical value of $5/4$, while exceeding the
maximum value permissible for a classical probability (namely $1$) by
\redlined{$13.32$ upper or $8.89$ lower percentile widths}. The directly observed quantity is an ordinary detector probability $p_D\le 1$ (or, with laser light, an equivalent power ratio); the value $\mu(E)>1$ is inferred via the calibrated relation $\mu(E)=2p_D$ for our filter.

If an unconventional theoretical concept is to play a role in meeting
the foundational challenges of quantum theory, it seems important to
bring it into contact with experiment as much as possible.  Our
experiment does this for the quantum measure.
\end{abstract}

\section{\label{sec:Intro}Introduction}

What exactly are we measuring when we perform a quantum measurement?
What, 
beyond the fact that a certain detector has clicked or the trace
on a certain oscilloscope has shown some specific behavior, 
does the measurement teach us about events in the physical world? 
A full answer to this question 
would go a long way toward resolving the foundational
puzzles raised by quantum theory 
because it would clarify the relationship that 
formal mathematical objects like wave functions and complex amplitudes 
bear to 
physical objects and events in the microscopic realm.\\

Consider for example an optical experiment 
involving an apparatus consisting of a laser together with components
such as mirrors, beam splitters, polarization optics, and photo
detectors mounted on an optical bench. 
Imagine now that a photon emerges from the laser and
makes its way through the various optical elements to a detector which
accordingly ``clicks'', and in the process destroys the photon, thereby
bringing the experiment to an end. What does the ``click'' of this
detector reveal to us about the micro-world?\\

Thanks to some combination of convention, habit, and prior experience,
we feel comfortable in saying that 
an observed
``click'' signifies the arrival
of the photon. One might question how this implied transformation of 
photon to 
observed 
click 
actually takes place (the so-called measurement problem); 
but even if we refrain from asking about this, 
a 
still
more glaring
explanatory gap remains. The final click tells us nothing definite about
the photon's career after it left the laser and before it arrived at the
detector. It fails to answer the question, 
\textit{What, exactly, happened in between the initial emission and the final detection}?\\

But if you believe (as almost any working physicist will) that something
really did happen in-between, then the paradigm of 
\phantom{$-$} preparation $-$ evolution $-$ observation \phantom{$-$} 
leaves much to be desired 
(not least because if we are honest, we have to admit that we
live our whole lives in this in-between condition). 
All the more interesting, therefore, are measurement procedures that aim to shed
light on entire physical processes and not just 
certain types of
correlation between an ``input'' and an ``output'' 
(or between a ``preparation'' and a ``detection'') $-$ 
procedures that concern themselves with temporally extended \textit{events} 
rather than just momentary \textit{states}. \\

\redlined{In this work, we realize an optical event filter.} From the standpoint of standard quantum optics, our apparatus implements a generalized measurement (\redlined{a positive operator-valued measure, POVM}) on the enlarged system that includes the photon's polarization as an ancilla. Accordingly, the raw experimental observable is an ordinary detector probability (or, with laser light, a power ratio) that is necessarily $\le 1$. \redlined{Here, the quantum measure $\mu(E)$ is a non-negative weight assigned to an event $E$ (a set of histories) that incorporates interference between histories and, unlike an ordinary probability, can exceed unity.} The novelty is not new dynamics, but engineering the measurement so that this probability equals a fixed fraction of the quantum measure of a chosen \emph{system event} $E$ in the underlying histories description. Strictly speaking, no ordinary probability exceeds unity here; what exceeds unity is the quantum measure $\mu(E)$ inferred from the detector probability via the event-filter mapping.\\

The experiment reported in the present paper implements a procedure of this sort.
It builds what we will call an \textit{event filter} 
that \textit{selects for a particular set $E$ of trajectories}, 
in close analogy to how a polaroid filter selects for a
particular polarization-state.  
Our experiment realizes in the laboratory 
(and for a particular event) 
a version of the general scheme described in \cite{Frauca_2017}.  
(Although it is inspired by \cite{Frauca_2017}, there are some important
differences, which will be noted in the sequel.)  The thing to be emphasized is
that what an event-filter measures cannot be described by a system-observable
that pertains to a single time: it ``filters for'' an event rather than a
momentary state. \\

By way of comparison with our experiment, one can mention two other examples of
event-oriented measurement schemes that address the question of
``What, exactly, happened in between'', 
namely
``negative result measurements'' \cite{NRM_Leggett_1988, NRM_Leggett_2008}
and
``weak measurements'' \cite{WM_AAV_1988}.\\

For an example of the former type of measurement, consider a photon which is
directed toward a polarizing beam splitter that transmits $H$ polarization and
reflects $V$.  If one then places a detector behind the beam splitter in the
transmitted direction, and if this detector \textit{does not click}, then 
(for an ideal scenario without absorption) 
one can conclude 
from this negative result
that the photon was reflected with $V$ polarization.  
In 
such a
measurement, 
the photon (in favorable runs) is not destroyed, 
and hence one has acquired ``in-between'' information on an ongoing physical process 
while allowing the process itself to continue.  
(We'll meet this technique again in Section \ref{sec:Discussion}, where we discuss an
enhancement of our experimental design that could be implemented in a future
iteration of our experiment.) \\

Weak measurements are also very relevant to our discussion, in view of their
intimate connection to path-integral formulations of quantum mechanics.
The output of a weak measurement \cite{WM_Duck_Sudarshan_1989},
is a complex number of the form
$\dfrac{\bra{f}\hat{O}\ket{i}}{\braket{f}{i}}$ ,
where $\ket{i}$ is the ``pre-selected state'' and $\ket{f}$ is the ``post-selected state'',
the complex number itself 
being known as the weak value 
$\ev{\hat{O}}_{w}$ of the operator $\hat{O}$ under the given experimental conditions.
This number, long known to quantum field theorists as an \textit{in-out expectation value}, 
is a transition amplitude of the sort that is ubiquitous in calculations of $S$-matrix elements; 
and it is directly reducible to a path integral.%
\footnote{When one thinks of $\ev{\hat{O}}_{w}$ as a transition amplitude (more
  correctly a ratio of them) rather than an expectation value, 
  one feels no astonishment that it fails in general to be a real number.}
Moreover this connection to path integrals (see also \cite{WeakValue_PathIntegral_2020}),
is no accident, we think, because \textit{histories-based} frameworks are the only ones
known to us that take as their starting point physical processes in their entirety,
rather than momentary states evolving temporally via the Schr{\"o}dinger
equation. As such they offer the only framework within which questions about
``what happened in between'' find a natural home. \\

In subsequent sections we will describe more fully our experiment and its
conceptual context, which is that of the histories-based framework known as
Quantum Measure Theory (QMT).
As a further development of path-integral formulations, 
QMT offers quantum mechanics a
formalism not founded on concepts like wave function, superposition of states,
or operators as observables \cite{QMT_Sorkin_1994}.
Instead it describes the kinematics (the ``ontology'')
of a physical system in terms of its \textit{histories} 
(e.g. a particle trajectory) 
and its \textit{events}, an event being a set of
histories.  
In line with mathematical formulations of the theory of stochastic processes
(from which the terminology ``event'' is also taken),
QMT describes the dynamics of the system in terms of a \textit{quantum measure}
that assigns to every event $E$ a generalized probability $\mu(E)$ 
(which, though non-negative, can exceed unity).\footnote%
{In the framework known as Decoherent Histories or Consistent Histories
\cite{Griffiths_stat, PhysRevLett.70.2201, hartle2014spacetimequantummechanicsquantum, PhysRevD.47.3345, Roland_logical,PhysRevA.55.4030}
  the same event $E$ would also receive a measure $\mu(E)$
  exceeding unity. In fact the computation would be exactly the same as
  here. The difference however is that in that framework $\mu(E)$ is
  regarded as meaningful only for events $E$ which decohere from their
  complements, i.e. events for which the measure of $E$ and the measure of
  its complementary event sum to $1$ (which of course is impossible when
  $\mu(E) > 1$). In contrast QMT treats every event as meaningful in
  principle, and our experiment supports this attitude by illustrating
  how the measure of an arbitrary event can acquire an operational
  meaning.}
\\

One of the simplest imaginable quantum systems is the ``two-site
hopper'' studied within the framework of quantum measure theory by
reference \cite{2-site_Hopper_Gudder2011}.  Our experiment realizes such
a hopper as a photonic system, 
and employs an ancilla-based filtering scheme (an event filter)
in order to determine the quantum measure $\mu(E)$ of a specified hopper event $E$. 
The experiment addresses a question which, posed in a classical setting,
would ask 
``Which path did the photon follow in going from source to detector?'' 
Pinning the photon down to a unique path, however, is
not a very enlightening exercise in a quantum context.
It is trivially accomplished by inserting detectors in suitable locations, but
in doing so, one suppresses the interference between distinct paths which is the hallmark of
quantum processes. 
Instead, we will select a 
\textit{set} of paths ---- an event ---
and ask the question,
``Did this event happen?''. 
An apparatus that implements this question in an appropriate sense is what we are calling an event-filter.\\

As already mentioned, the specific event filter used in our experiment contains
some novel features with respect to the class of protocols set out in \cite{Frauca_2017}.
Firstly
(as discussed in detail in Section \ref{sec:Experiment})
we employ the photon's polarization degree of freedom as our ancilla;
and secondly
(since no additional ancilla is available) 
we are led by necessity to \textit{reuse} ours 
by in effect coupling it a second time to the system.
%
The use of the photon's polarization as ancilla 
(rather than some totally separate ``qubit'') 
is merely an artifice that allows us the convenience of employing
polarizing beam-splitters to couple the ancilla to the photon's trajectory.
But the technique of coupling the same ancilla to the system more than once
offers an improvement over \cite{Frauca_2017} that could be important in
application to more complicated events containing large numbers of histories.
(In this simple case, it raises the counting efficiency from 1/3 to 1/2.) \\

The fact that our event-filter yields correctly the theoretically computed
value of $\mu(E)$ is strong evidence that it is
functioning correctly.  In our present design, though, the photon gets destroyed
by a detector placed at the desired output port, thereby bringing the experiment
to an abrupt end.
In that sense, our design falls short of what one would expect from a true
filter, which would pass the photon through without destroying it.
Enhancing our design to achieve this behaviour would not be very difficult, however.
As described in Section \ref{sec:Discussion}, a simple negative-result
modification of our apparatus  would render the filter non-destructive, allowing
the experiment to continue. 
Subsequent testing
employing further optical elements 
could then verify that the filter had truly
selected for the intended trajectories.\\

In connection with quantum computing, a fully non-destructive event-filter would
also have potential applications that are not hard to imagine.\\

In the remainder of this paper, we first provide some background that is
helpful to appreciate more fully the significance of our experiment, and then we describe the
experiment itself, its results, and some of its implications as we see
them. 
The background and an overview of the history-based framework of QMT
are presented in section \ref{sec:QM_to_QMT},
followed in section \ref{sec:simplified} by a simplified version of our experiment intended to
illustrate the main ideas behind it.
The experimental details and the results are
described in Sections \ref{sec:Experiment} and \ref{sec:Results},
with
sections \ref{sec:Discussion} and \ref{sec:Conclusion} presenting 
final comments, interpretations, and proposals for further experiments
that would enhance or extend the one we have performed.\\

\section{\label{sec:QM_to_QMT}Overview: Conventional Quantum Theory to Quantum Measure Theory}

In what one might refer to as the standard quantum formalism, a
microsystem is described by a wave function $\psi$ that resides within a
Hilbert space, and the dynamics of the system is governed by 
a
Schr\"{o}dinger equation that generates unitary time-evolution of $\psi$
under a particular Hamiltonian. A property of the system manifests
itself in the outcome of a measurement of an observable
$\hat{O}$. 
%
The probability distribution over different possible outcomes can be
obtained from $\psi$ and $\hat{O}$ by using the Born rule
\cite{BornRule_Landsman2009}.  Although the predictions made by this
formalism have always been found to be consistent with experiments, the
theory appears to be inadequate as it fails to comment on the reality of
the micro-system at intermediate times between preparation and
observation. Nor does it address the question of what exactly takes
place during the measurements (or preparations) themselves \footnote{A
  1962 talk by Wendell Furry \cite{Furry_talk} offers an exceptionally
  clear account of this second aspect. We limit ourselves to quoting
  just the following extract. 
  ``The important thing is, the statement is simply: 
  \textit{when you measure, this is what you get}. There is
  \textit{no statement made} as to what happens in the actual measuring
  process. \dots This is what various people \dots call the cut. It
  is where something happens which the theory does not describe
  mathematically.''}.
These two lacunae lie at the core of the so-called quantum measurement
problem, and have given rise to continuing debates over whether one can
assign an ontological meaning to $\psi$ \cite{Ontology_of_Wavefunction},
and how to conceive of what (if anything) happens between preparation
and measurement. \\

Amidst all these debates and attempts to provide a complete description
of the quantum system and its behaviors, one finds oneself continually
returning to the fact that the conventional interpretation, with its
emphasis on the measurement-process, 
presupposes an a priori division of
the universe into an observer (classical system) and an observed
(quantum system); but it does so without providing any coherent account
of this ``Heisenberg's cut'' \cite{Heisenberg_Cut_Zurek_1991, Quantum-Classical_Cut_Ellis_2012}. 
This has led many people to
feel the need for a more unified conception that could encompass both the micro and macro worlds within a single framework.\\

In this quest, one can observe a contrast between approaches that have
inherited the pre-relativistic notion of momentary states evolving in
time, and approaches that adopt a more global, spacetime point of view
from the very beginning. 
An approach of the former type will typically
conceive of $\psi$ as a function defined on configuration-space and
evolving continuously over time. In contrast, manifestly covariant
descriptions are more natural to gravitational physics, as well as to
high energy processes and quantum field theory in Minkowski spacetime.
For these reasons, it has been argued that the Dirac-Feynman path
integral \cite{Feynman_1948}, or more generally a sum-over-histories
approach towards quantum 
dynamics like that in \cite{SOH_EPRB_Sinha1991},
is more appropriate in adapting quantum mechanics to special relativity,
and especially to general relativity
\cite{SpaceTime_Approach_to_QM_Hartle_1993}. \footnote 
{This allies with the philosophy of ``gravitizing quantum mechanics''
\cite{Penrose_2014}, i.e. modifying it to fit better with general
relativity, as opposed to only ``quantizing gravity'', i.e. modifying
general relativity to fit better with existing formulations of quantum
mechanics and quantum field theory
\cite{Decoherence_Functional_Linden1994}. }\\

Moreover, a formalism that, like Quantum Measure Theory (QMT)
\cite{Sum_Over_Histories_in_Gravity_Sorkin, QMT_Interpretation_Sorkin_1997}, 
could dispense 
at a fundamental level with concepts like wave function, observable, and
state-vector reduction, would be better suited to cosmology and the
study of the early universe \cite{Role_of_Time_in_QM_Hartle1988}.  This
is because on the one hand, quantum effects cannot be neglected in the
early universe, while on the other hand, and unlike in a laboratory
setting, nothing in the physics of that epoch obviously corresponds to
the concepts of observer or state preparation and measurement. 

\subsection{\label{subsec:QMT}Quantum Measure Theory: A Histories-based Formulation}

Quantum Measure Theory (QMT) offers a spacetime formulation of quantum
mechanics based on a path-integral or sum-over-histories approach. It
interprets the behavior of a quantum system from the perspective of a
suitably generalized theory of stochastic processes
\cite{QMT_Sorkin_1994}. The kinematics or ``ontology''
of such a theory rests on the concepts of \textit{history} and
\textit{event}; and the dynamics is a kind of quantum-stochastic law of
motion for the history \cite{SOH_EPRB_Sinha1991}, given mathematically
by a \textit{quantum measure} which generalizes the classical notion of
a probability-measure so as to incorporate quantum interference. \\

A \textit{history} in this framework is taken to be the fundamental
building block of reality; it gives the finest grained description
conceivable in the theory of the system in question. If, for example,
the system in question were a single particle, a history would be a
single worldline; if it were a field, a history would be any of its
conceivable spacetime configurations, etc.  In general, the definition
of history depends on the system one is dealing with and the model or
context adopted to describe it. 
An \textit{event} is then identified
with a set of histories of the system, and it is mapped to a
non-negative real number, called its \textit{quantum measure}, that
generalizes the classical notion of the probability of a random
event.\footnote{Notice that when one speaks of ``the event $E$'' one
  does not imply that $E$ really happens, only that it might or might
  not happen in any particular case. For this reason David Reid has
  suggested the alternative term ``occurable'' instead of ``event''.
  However, for now we stick with the usage that has become standard in
  probability theory.}
%
Formulated in this manner, a quantum theory is more akin to a classical
theory of stochastic processes like Brownian motion than it is to
classical Hamiltonian dynamics.\\


In quantum measure theory, the kinematics and dynamics of a given system
are defined mathematically by a triple $(\Omega, \mathcal{A}, \mu)$;
where $\Omega$ (the history space) is the space of all conceivable
histories, $\mathcal{A}$ (the event algebra) is the set of all subsets
of $\Omega$ to which a measure can be assigned (including the empty set
$\emptyset$ and $\Omega$ itself), and $\mu$ (the quantum measure) is a
function that maps each element of event algebra to a positive real
number ($\mu: \mathcal{A} \to \mathds{R}^{+}$). Unlike a classical
probability, a quantum measure incorporates quantum interference, and
hence, its values cannot (except in important special cases) be
interpreted as probabilities in the usual sense. They neither obey the
probability sum rule nor (because of constructive interference) are they
bounded above by unity. \\


The quantum measure $\mu$ can be characterized abstractly by certain
positivity conditions and axioms that generalize the Kolmogorov sum rule
for probabilities. For a large class of theories $\mu$ takes the form of
a ``double path integral'' designed in such a way that when $E$ is what
we will term an ``instrument event'', $\mu(E)$ will 
be equal to 
the
Born-rule probability of the instrument ``reading'' corresponding to
$E$. 
For the type of unitary system that figures in our experiment,
where a history corresponds to a particle-trajectory $\gamma$ restricted
to a definite interval of time, and for an event $E = \{ \gamma^{1},
\gamma^{2}, ....\}$ comprising a finite number of histories, $\mu(E)$ is
given by the formula,
\begin{align}
    \mu(E) = \sum_{\gamma^{i}, \gamma^{j} \in E} A(\gamma^{i}) A^{*}(\gamma^{j}) \delta_{\gamma^{i}_{end} , \gamma^{j}_{end}}
\label{eqn:Measure}
\end{align}
\noindent
Here, $A(\gamma)$ is (as illustrated below) the quantum amplitude
associated with the history $\gamma$, and the delta-function
$\delta_{\gamma^{i}_{end}, \gamma^{j}_{end}}$ limits the interference
between histories to those that terminate at the same point. 
Because of
how the measure of an instrument event relates to the Born rule, the
measure $\mu$ for a given system encodes in particular all the
predictions about that system made by the ordinary quantum formalism. \\

\subsection{\label{subsec:Events}Different types of Events and their Measures}
Events, as defined in QMT, can be broadly classified into two
categories, instrument events and non-instrument events (one might also
name the latter as ``system events'').\\

Instrument events concern the histories of an instrument; by instrument
here we mean a piece of 
measuring apparatus, and the intended histories are those of the
variables that describe the instrument's macroscopic behaviour, in
practice the variables that express the ``output'' of the
measurement. For example, a particle detector (in our experiment, a
photodetector) could have the history space, 
$\Omega_{I} \equiv \{\checkmark, \cross\}$;
%
%
where $\checkmark$ and $\cross$ are the two histories of the detector
corresponding respectively to the two classically admissible outputs,
`click' and `not-click'. In this case, the event algebra contains 
four instrument events 
$\mathcal{A}_{I} = \{\emptyset, \{\checkmark\}, \{\cross\}, \{\checkmark, \cross\} \}$, 
with the measures of the first and last events being $0$ and $1$, respectively. The measure
$\mu_{\checkmark}$ of the event $\{\checkmark\}$  gives the probability of detection, 
and we have $\mu_{\checkmark} + \mu_{\cross} = 1$, provided that the detector functions perfectly.\\

Non-instrument events, on the other hand, are events that do not involve
any instrument (specifically any detector). For example, the photonic
events associated with a photon encountering a lossless optical beam
splitter are such events, and the history space corresponding to this
episode in the photon's career can be taken to be $\Omega_{NI} \equiv
\{\mathcal{T}, \mathcal{R}\}$, where $\mathcal{T}$ and $\mathcal{R}$ are
the trajectories of the photon undergoing transmission or reflection
through the BS, respectively. 
Non-instrument events can be made more and
more complex by introducing more devices with multiple output ports in
the path of the system.\\

The classification into instrument and system events is not intended to
be either precise or exhaustive.  For one thing, microscopic events
happen within instruments as well as outside them, but such events are
of secondary interest as long as the functioning of the instrument can
be taken for granted, being deemed not to be in need of a detailed
analysis. \\

Although system events are not directly observed, they can often be
inferred from suitable instrument events. Whether a photon was
transmitted or reflected from an optical beam splitter remains unknown
to us until a detector registers the photon either in the transmitting
port or in the reflecting port of the beam splitter. It's well to
recall, however, that this kind of deduction is only justified to the
extent that the detector has high efficiency, because a dark count in the detector can falsely imply the presence of
the photon. Deducing the occurrence of a non-instrument event from an
instrument event relies on the assumption that there exists a perfect
correlation between the two classes of events, an assumption that needs
to be substantiated either by calibrating the instrument empirically, or
by analyzing its functioning theoretically when this is possible. \\


A further distinction between two different categories of
system events is important in connection with our experimental protocol.
This is the distinction between
\textit{serial} events and \textit{non-serial} events.  
Serial events include all those that could in principle be directly
identified from a \textit{sequence} of momentary instrument events,
without the need for any quantum ancilla. 
More abstractly a serial event is one that corresponds mathematically
with a sequence of projection operators, and thereby with a
corresponding sequence of projective ideal measurements.\footnote%
{It follows in particular that the quantum measure of such an event can never
  exceed unity.}\\

In contrast, a non-serial event $E$ is one whose occurrence cannot be revealed
in this simple manner.
Detection of such an event seems to require the use of quantum ancillas which
can gather information about the system's history (for us the path of the photon),
in a more controlled manner.
Ideally, they would gather no information beyond the bare amount needed
to affirm that ``The event $E$ has happened'', but when this is impractical,
the ancillas can be made to interact with each other subsequently, 
so as to forget whatever superfluous information they might have retained.
For example, the universal scheme of \cite{Frauca_2017} functions this way.%
\footnote {This suggests a notion of complexity for system-events such that one event is
   more complex than another if a filter for it requires more ancillas or 
   requires an ancilla to couple more times to the system.}
From this perspective, the event-filter of our experiment can be understood as
using a single ancilla that couples twice to the system. \\

Operationally, this distinction matters because in an ordinary interferometer a detection at a single output port corresponds to a \emph{serial} (instrument) event, whose measure reduces to an ordinary probability and therefore cannot exceed unity. What the event-filter adds is a controlled use of an ancilla to \emph{convert} a chosen non-serial \emph{system} event $E$ into a single detector event $D$, so that the detector probability $P(D)\le 1$ encodes $\mu(E)$ up to a known, device-determined factor.\\

A practical example might help to clarify the distinction between serial 
and non-serial events.

Consider a particle passing through two double slit diaphragms, one
placed after the other, separated by a finite gap. Let $A_{1}, B_{1}$
represent the upper and lower slits in the first diaphragm, and $A_{2},
B_{2}$ represent the same in the second diaphragm.
The events, 
$E_{1} = \{A_{1}A_{2}, B_{1}A_{2}\}$ 
and 
$E_{2} = \{A_{1}A_{2}, A_{1}B_{2}\}$, 
are then to be considered as serial events, as they can
be detected simply by a placing an instrument (say a detecting screen)
after the second diaphragm, while blocking the slit $B_{2}$ (for
$E_{1}$) or $B_{1}$ (for $E_{2}$). 
In words, $E_{1}$ is the event that
the photon traverses ``either of the two slits in the first diaphragm
\textit{and then} the upper slit in the second diaphragm'', and
similarly for $E_{2}$. Even more simply the event $E_{3} =
\{A_{1}B_{2}\}$ is also a serial event, because it answers to the phrase
``upper slit in the first diaphragm \textit{and then} lower slit in the
second diaphragm''. 
On the other hand the events, $E_{4} = \{A_{1}A_{2},
B_{1}B_{2}\}$ and $E_{5} = \{A_{1}B_{2}, B_{1}A_{2}\}$, are not serial
events, as they cannot be detected by masking any subset of the slits
and then asking whether or not the photon has arrived at the final
screen. \\

Serial events clearly form a very special category within the full
event algebra of a system.  The possibility to design event filters
corresponding to system events which are not serial opens up therefore a
much greater range of possible measurements than is ordinarily
contemplated. By dealing with all kinds of events, including non-serial
events, quantum measure theory goes beyond ordinary QM in the direction
where a better understanding of relativistic quantum field theory, and
beyond that of quantum gravity, appear to lie.\\

\paragraph{Roadmap and notation.}
To separate conceptual aspects from experimental details, we first present a pedagogical example that illustrates how a quantum measure is computed for a chosen
(non-serial) event. The specific event used in this pedagogical section, denoted
$E_{\rm demo}$, is selected only for clarity and is \emph{not} the event implemented in our
experiment. The experimental realization is described in Section 4, where the event of
interest is $E_{\rm exp}=\{00,01,11\}$. To avoid confusion, we will explicitly distinguish
$E_{\rm demo}$ discussed in section 3 from $E_{\rm exp}$ (discussed in Sections 4 and 5).\\

\section{\label{sec:simplified}A simplified experiment that illustrates the main ideas}

\noindent\textbf{Pedagogical example (not the experimental event).}
In this section, we consider a simplified illustrative event $E_{\rm demo}$ to explain the
basic definitions and the structure of the interference terms in the quantum measure.
This example is intended purely as a conceptual guide; the experimentally implemented
event and its corresponding event-filter are introduced later in Section 4. All expressions derived here are general and apply to any event once the corresponding
histories (and intensities/phases) are specified.
\\

Let us consider a simplified design intended to bring out the main ideas behind the experiment. In doing so, it is important to distinguish mentally the pure ``System''
itself, and events therein, from the larger experimental setup that
includes as well those additional optical elements that help to separate
out the System-event $E_{\rm demo}$ for which the filter has been designed, and
whose measure $\mu_{\rm demo}(E_{\rm demo})$ we wish to measure.\footnote%
{We intend these ``additional elements'' to include the photon's
 polarization, which one is separating conceptually from its
 trajectory or ``worldline''.}
Correspondingly two different quantum measures come into consideration,
one for the System per se, and one for the full experimental setup
including all the components of our event-filter.  In this section we
will use the symbol $\mu_{\rm demo}$ for the former and $\mu_{F,\rm demo}$ for the latter.\\

\begin{figure*}[t]
    \centering
    \includegraphics[width = 0.65\linewidth]{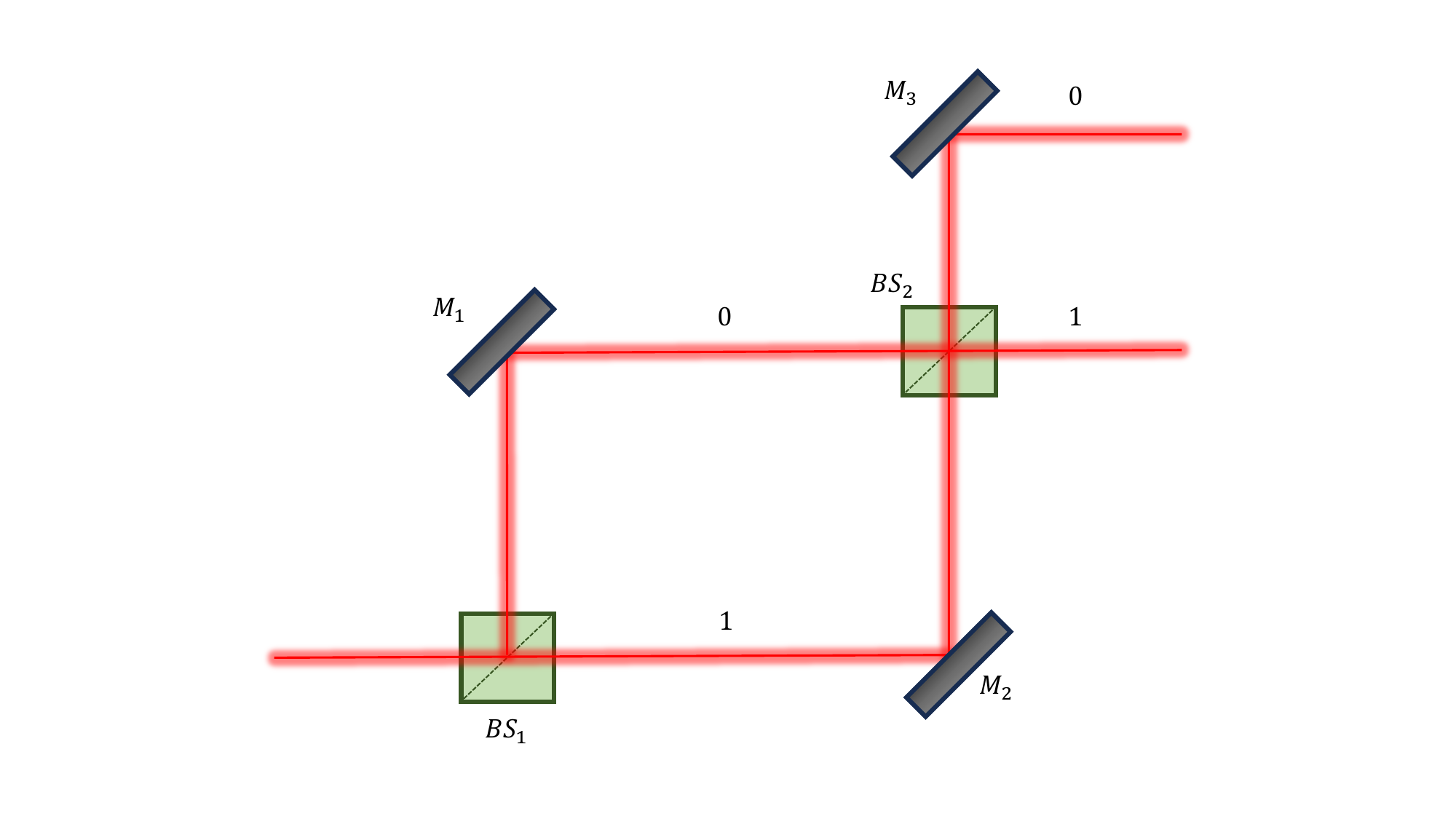}
    \caption{Schematic of the photonic \emph{system} (not the event-filter): a photon propagates through beam splitters $BS_{1}$ and $BS_{2}$ and mirrors $M_{1}$--$M_{3}$, defining the four two-step histories $\Omega=\{00,01,10,11\}$ (with $0/1$ denoting the upper/lower output mode after each beam splitter) used in the pedagogical discussion of Section 3. The experimental event-filter is introduced in Section 4.}
    \label{fig:system}
\end{figure*}

Our {\it\/system\/} is depicted schematically in Figure  \ref{fig:system}.
It comprises a photon (more correctly, only its position), together with a set
of mirrors and beam-splitters arranged in the form of a Mach-Zehnder Interferometer.  
For simplicity we limit ourselves to two beam splitters, although one could
continue the pattern indefinitely.  
If we exclude the optical elements from our history-space, as we may do since
they remain the same from history to history, then a system-history is
determined by which of the two output ports the photon follows in each
beam-splitter.
The upper or lower path taken by the photon after each
beam splitter will be labelled as 0 or 1 respectively. 
Thus, the history space for this System can be idealized as $\Omega = \{00,01,10,11\}$. \\

The quantum amplitudes associated with passage through a 50:50 beam splitter can always be taken to be 
$1/\sqrt{2}$ for transmission and $i/\sqrt{2}$ for reflection.
 The net amplitudes for the four histories in $\Omega$ will then be
$A(00)=-1/2$,
$A(01)=i/2$,
$A(10)=+1/2$,
$A(11)=+i/2$,
where we have ignored propagation amplitudes that are common to all the
paths.\\

In order to illustrate how the quantum measure is defined, we can
consider three possible System-events in this pedagogical setting, the first of which is just the
singleton-event $E_{\rm demo}^{(1)}=\{00\}$.  For this event, 
equation (\ref{eqn:Measure}) yields simply
$\mu_{\rm demo}\!\left(E_{\rm demo}^{(1)}\right)=|A(00)|^2=1/4$, 
which is recognizable as the probability for the
photon to activate a detector located after mirror $M_3$.\\

Our second event will be $E_{\rm demo}^{(2)}=\{00,10\}$.  Since both of these paths end at the
same point, the $\delta$ factor in (\ref{eqn:Measure}) equals unity, and the measure of $E_{\rm demo}^{(2)}$
becomes 
$|A(00)+A(10)|^2 = |-1/2 + 1/2|^2 = 0$.  This vanishing of the
quantum measure of $E_{\rm demo}^{(2)}$ is recognizable as 
the familiar fact that the
output beam 0 is dark in a symmetric MZI.\\

Thirdly, consider an event similar to the one to which our experiment is devoted, namely
$E_{\rm demo}=\{10,01,11\}$.  
In this case  (as per equation (\ref{eqn:Measure})) only histories $01$ and $11$ interfere,
and we find
$\mu_{\rm demo}(E_{\rm demo}) = |i/2+i/2|^2 + |1/2|^2 = 5/4 = 1.25.$ 
Unlike our first two examples, this 
is a non-serial event, 
and its measure has no immediate interpretation as the probability of anything.
But by designing an event-filter for $E_{\rm demo}$, we will be able to interpret
$\mu_{\rm demo}(E_{\rm demo})$ as twice the probability $P$ that the filter will register a positive outcome: $P=\mu_{\rm demo}(E_{\rm demo})/2$.  \\
 \\

Just as in \cite{Frauca_2017} 
\footnote%
{which however would have produced a factor of 3, as mentioned earlier. Our experimental design improves this to a factor of 2.},
this factor of 2 is a property of the filter and not of the system per se.
It would not
change if one were to insert phase shifters into any of the beams, or if
the transmission and reflection coefficients of the beam splitters were
other than $1/\sqrt{2}$.  
It is this ``universality'' that justifies our claim that by means of our
filter we are able to ``measure the quantum measure of $E_{\rm demo}$''.
\bigskip
\begin{figure*}[t]
    \centering
    \includegraphics[width = 0.65\linewidth]{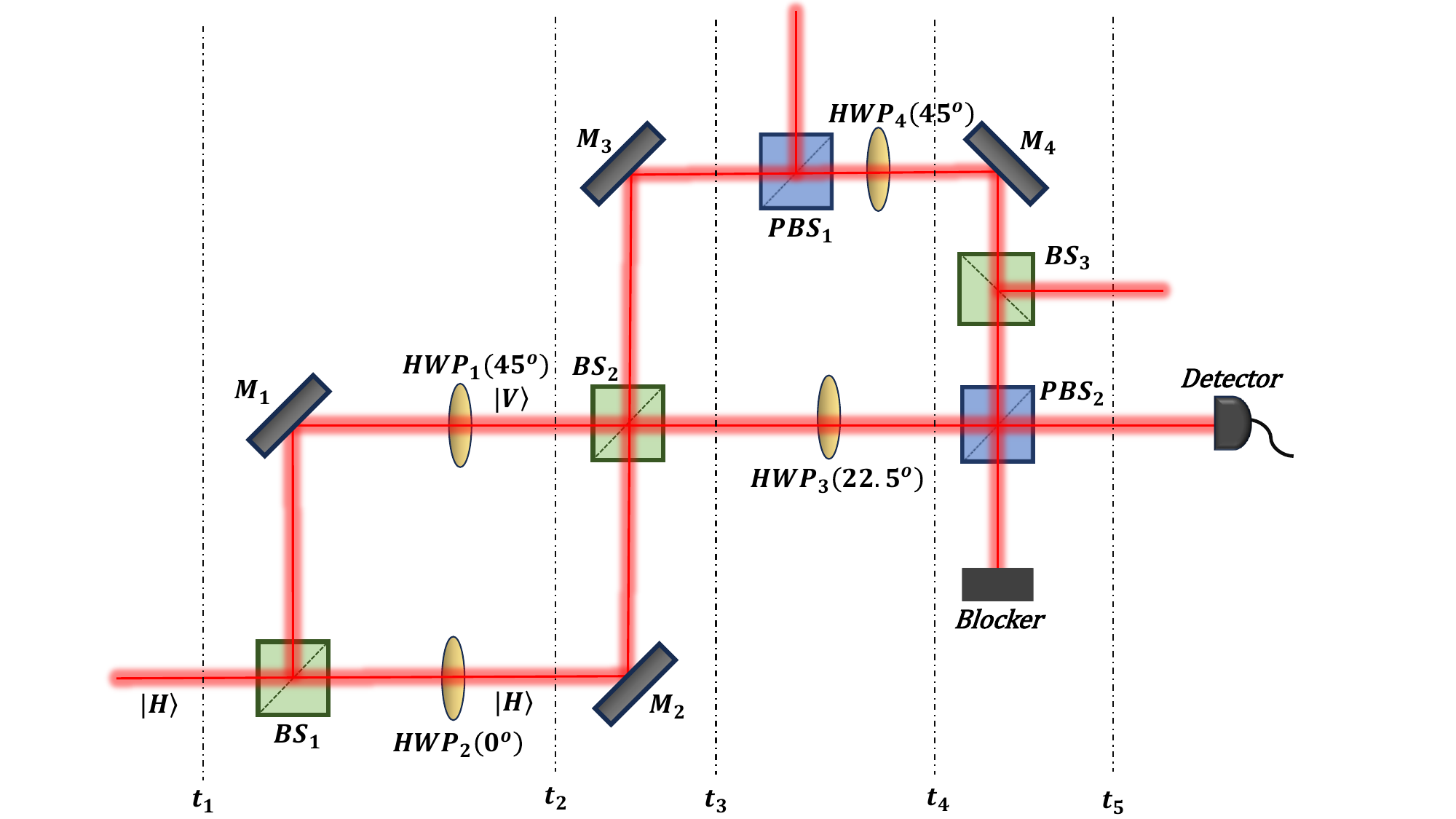}
    \caption{Pedagogical event-filter for the illustrative non-serial event $E_{\rm demo}=\{01,10,11\}$, drawn as an ``unfolded'' version of the displaced Sagnac geometry to make the filter logic explicit. The system elements (beam splitters $BS_{1},BS_{2}$ and mirrors $M_{1}$--$M_{3}$) are the same as in Fig.~1; the additional polarization optics together with the detector constitute the filter. For clarity, the excluded history is shown as $00$ (removed by $PBS_{1}$) in this illustrative design. The experimentally implemented event-filter (Sec.~IV), corresponding to $E_{\rm exp}=\{00,01,11\}$ and rejecting history $10$, is shown in Fig.~4.}
    \label{fig:simplified}
\end{figure*} 
\bigskip

A specific filter for the event $E_{\rm demo}=\{01,10,11\}$ is shown in Figure \ref{fig:simplified}, 
which ``unfolds'' the arrangement of Figure \ref{fig:QMT_setup} below 
to show 
the underlying design more
perspicuously. 
However, note that while in the actual experiment history
10 is rejected, here we illustrate the rejection of 00. 
We do this purposely to demonstrate that indeed many 
combinations of histories
are possible that result in non-serial events. 
Because the polarization
degree of 
freedom of the photon is being recruited as an ancilla in the
sense of \cite{Frauca_2017}, 
the distinction between System and Filter is somewhat blurred, 
but with a little effort, and referring to Figure \ref{fig:system}, 
one can still recognize the beam splitters $BS_1$, $BS_2$ 
and the mirrors $M_1$, $M_2$, and $M_3$, as belonging to the System,
while the other optical components together with the detector and the
photonic polarization itself can be recognized as belonging to the Filter.  
Notable are the half-wave plates (HWP) that couple the photon's polarization to its trajectory,
the polarizing beam splitter $PBS_1$ that expels the excluded history $00$, 
and the $PBS_2$ that undoes the way that the polarization was used
to tag certain of the histories.  
(Its companion $BS_3$ is needed only to balance the complementary outputs associated with the $HWP_2$--$PBS_2$ stage in this unfolded schematic; it should not be interpreted as physical absorption or ``loss'' in $PBS_2$.)\\

Let $D$ be the event that the photon 
reaches 
the detector.  One can
compute $\mu_{F,\rm demo}(D)$ similarly to how the measures $\mu_{\rm demo}$ were computed in
the above examples, and one finds that $\mu_{F,\rm demo}(D) = \mu_{\rm demo}(E_{\rm demo})/2$.  If
one seeks the origin of the 1/2 in this equation, one can locate it in the fact that the final $HWP_2$--$PBS_2$ stage produces two complementary outcomes in the ancilla basis and we condition (postselect) on the detector outcome $D$; in the ideal symmetric setting this postselection has success probability $1/2$.
This corroborates our earlier assertion that the factor of 1/2 belongs to
the filter and not to the system.\\

\noindent\textbf{Transition to the experiment.}
Having established the notation and the structure of $\mu_{\rm demo}(E_{\rm demo})$ using the illustrative event
$E_{\rm demo}$, we now return to the experimental implementation. In Section 4, the event of
interest is $E_{\rm exp}=\{00,01,11\}$, and all subsequent references to the event-filter,
recorded powers, and experimentally inferred measures pertain to this experimental event.

\section{\label{sec:Experiment}Experimental Determination of the Quantum Measure of a Photonic Event}  

\noindent\textbf{Experimental event and event-filter.}
We now describe the experimental realization of the event-filter corresponding to the
non-serial event $E_{\rm exp}=\{00,01,11\}$. The optical network is designed such that the
histories in $E_{\rm exp}$ interfere at the monitored output, enabling an operational
estimation of the associated quantum measure. Henceforth in Sections 4 and 5 we write $E\equiv E_{\rm exp}$ for brevity.\\

For easy reference, we redisplay here the essential portion of Figure \ref{fig:system},
turned sideways in order to exhibit its inherent symmetry more clearly 
(and also to bring out how it realizes the first two stages of the 2-site-hopper model). 

\begin{figure*}[t]
    \centering
    \includegraphics[width = 0.65\linewidth]{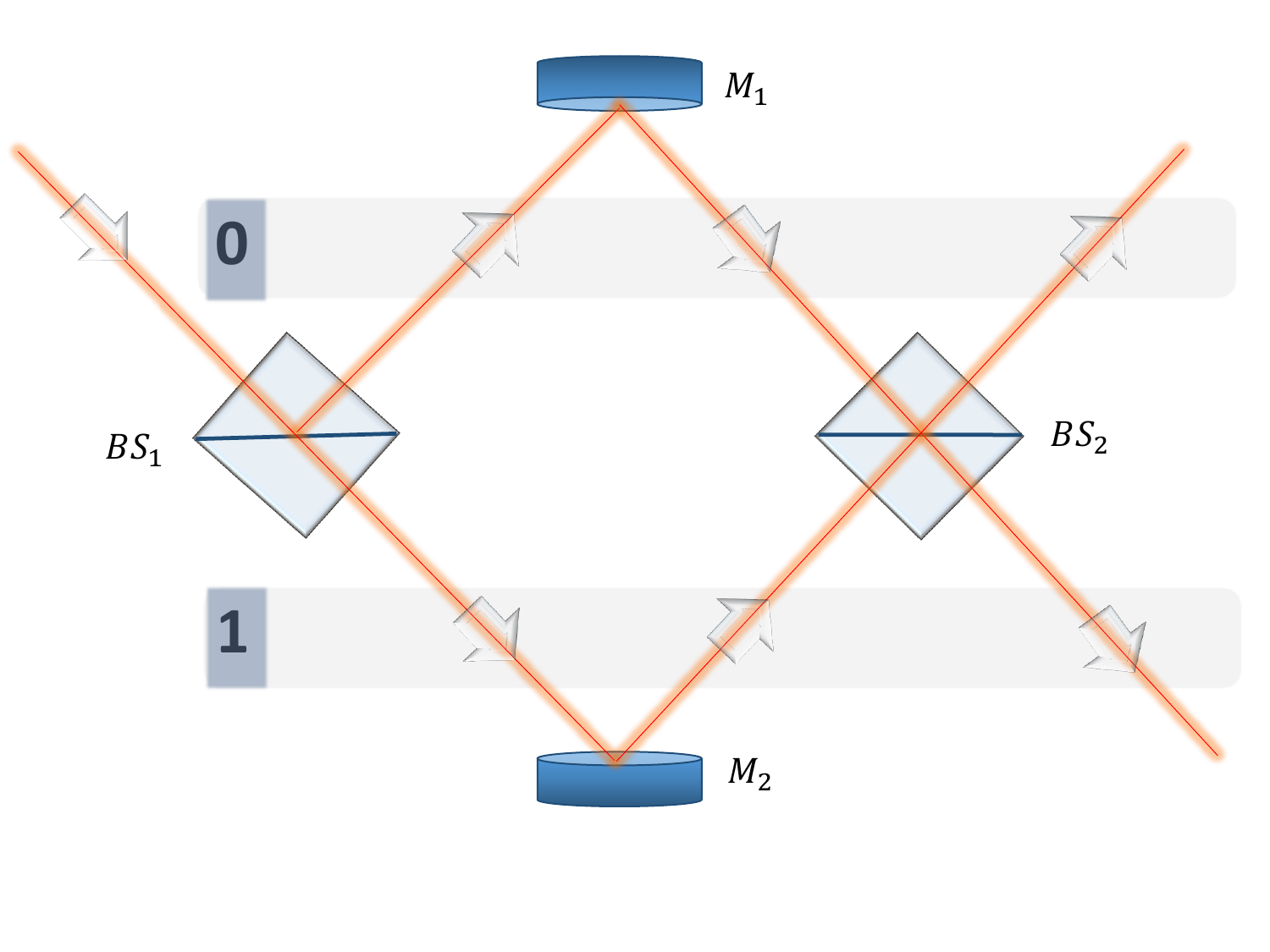}
    \caption{Conceptual schematic of the \emph{system} (not the event-filter) used for the analytic description in Section 4: a photon propagates through two beam splitters $BS_{1}(t_{1},r_{1},\phi_{1})$ and $BS_{2}(t_{2},r_{2},\phi_{2})$, with mirrors $M_{1}$ and $M_{2}$ redirecting the outputs of $BS_{1}$ into the inputs of $BS_{2}$, thereby defining the four two-step histories $\Omega=\{00,01,10,11\}$ (with $0/1$ denoting the upper/lower output mode after each beam splitter, as marked). This Mach--Zehnder representation is shown for clarity; the actual experimental implementation uses the displaced-Sagnac geometry and the full event-filter (including polarization optics and detection ports) is shown in Fig.~4. If iterated indefinitely to the right, this system would become an optical implementation of the two-site-hopper discussed in Refs.~[14,1].}

    \label{fig:Schematic}
\end{figure*}

Recall that by \textit{the system} we refer to the photon propagating within the
arrangement shown in figure \ref{fig:Schematic}.

\bigskip

\begin{figure*}[t]
	\centering
    \includegraphics[width=0.99\linewidth]{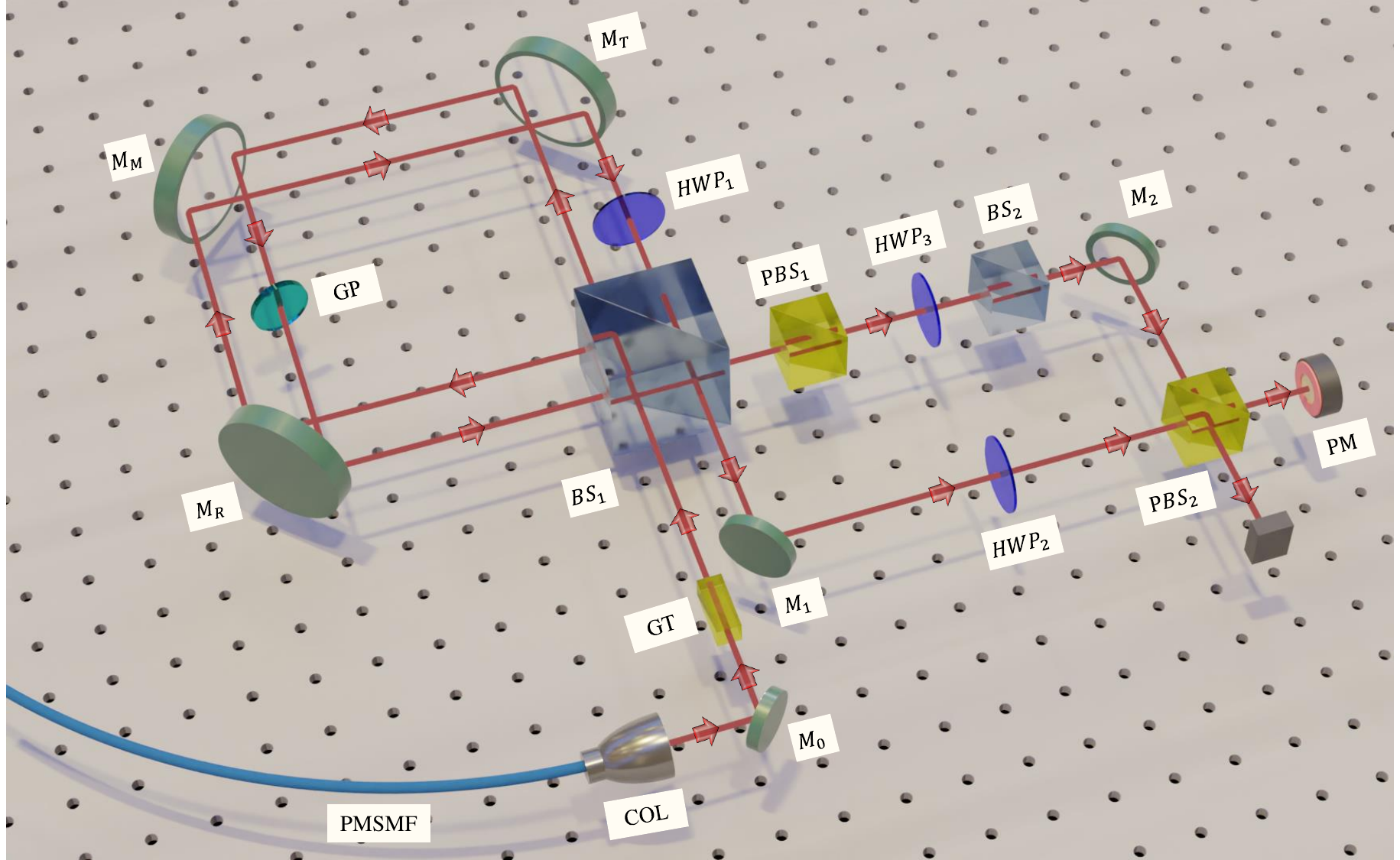}
\caption{Experimental implementation (Section 4) of the \emph{event-filter} for the non-serial event $E_{\rm exp}=\{00,01,11\}$ (the event measured in this work). The \emph{system} is the displaced Sagnac interferometer (DSI), in which the beam traverses the same beam splitter $BS_{1}$ twice, thereby defining the four two-step histories $\Omega=\{00,01,10,11\}$; the event-filter is designed to \emph{accept} the histories in $E_{\rm exp}$ and \emph{reject} the excluded history $10$. The \emph{filter} consists of the polarization optics and output ports (e.g., HWP/PBS/auxiliary optics and \redlined{the output monitored by the power meter (PM)}), which use the photon’s polarization as an ancilla to tag, recombine, and postselect the desired histories. The quantum measure $\mu(E_{\rm exp})$ is inferred from the output-to-input intensity ratio recorded at the filter ports (see Section 5). For conceptual clarity, the pedagogical “unfolded” schematic in Fig.~2 corresponds to an illustrative event $E_{\rm demo}$ and is not the experimental event implemented here.}

	\label{fig:QMT_setup}
\end{figure*}

The mirrors ($M_{1}, M_{2}$) and beamsplitters ($BS_{1}, BS_{2}$)
seen 
in figures \ref{fig:Schematic} and \ref{fig:system} 
form a Mach-Zehnder Interferometer (MZI).
However, for technical reasons, 
our actual experiment used 
the Sagnac configuration (DSI) shown in figure \ref{fig:QMT_setup}.
Since 
the simpler MZI configuration
is conceptually equivalent to 
the DSI configuration, 
we will remain with it for the short calculation that follows.\\

A photon incident on a (lossless)
beamsplitter emerges in either the transmitted or reflected beam.
The history space for our system (the photon) can thus be taken to be 
$\Omega^{(2,2)} = \{\mathcal{T}_{1}\mathcal{T}_{2}, \mathcal{T}_{1}\mathcal{R}_{2},
\mathcal{R}_{1}\mathcal{T}_{2}, \mathcal{R}_{1}\mathcal{R}_{2} \}$, 
where $\mathcal{R}_{1}\mathcal{T}_{2}$ represents the path of the photon undergoing
reflection in first $BS$ and transmission in second $BS$, etc.  
We will express these 4 histories in the simpler form used earlier:
$\Omega = \{00,01,10,11\}$, 
where the upper and lower spatial modes of the photon after a $BS$
are labelled as $0$ and $1$ respectively, as marked in Fig. \ref{fig:Schematic}.

\paragraph{Aim:} The aim of the experiment is to infer the value of 
 the \textit{quantum measure} $\mu(E)$ for the event $E=\{00,01,11\}$, thereby
 illustrating how a non-serial event can be detected experimentally, 
 and how the value of its quantum measure can be given an experimental meaning, even when it cannot be
 interpreted as a probability. \\

\paragraph{Computing $\mu(E)$ in an ideal scenario:}
Let, $t_{i}$ and $r_{i}$ respectively represent the transmission and reflection
coefficients of $i$-th beamsplitter $BS_{i}$, which is assumed to be lossless,
giving $\abs{t_{i}}^{2}+\abs{r_{i}}^{2} = 1$. 
Assuming that $t_{i},r_{i}\in\mathds{R}$ and that a phase $\varphi_{i}$ is acquired by the photon upon
reflection, we obtain the amplitudes associated with the possible histories as
$A(00) = r_{1}e^{i\varphi_{1}}r_{2}e^{i\varphi_{2}},\ 
 A(01) = r_{1}e^{i\varphi_{1}}t_{2}, \ 
 A(10) = t_{1}t_{2}, \ 
 A(11) = t_{1}r_{2}e^{i\varphi_{2}}$. 
Hence, from Eqn. \ref{eqn:Measure}, the quantum
measure $\mu(E)$ for the event $E=\{00,01,11\}$ is found to be,  
\begin{align}
	\mu(E) &=  \abs{A(00)}^{2} + \abs{A(01) + A(11)}^{2}
	\label{eqn:mu_E_formula}\\
	\mu(E) &= \abs{r_{1}r_{2} e^{i\varphi_{1}}e^{i\varphi_{2}}}^{2} \nonumber\\
	&\quad + \abs{r_{1}t_{2}e^{i\varphi_{1}} + t_{1}r_{2}e^{i\varphi_{2}}}^{2}
\end{align}
\noindent
Specializing to symmetric lossless $50:50$ beamsplitters, $t_{i} = r_{i} = \dfrac{1}{\sqrt{2}}$ and $\varphi_{i} = \dfrac{\pi}{2}$, 
we obtain for $\mu(E)$ 
the same value obtained earlier for the event $\{10,01,11\}$ i.e. $E_{\rm demo}$ of Section 3:
\begin{equation}
\begin{split}
\mu(E) &= \abs{ \dfrac{i}{\sqrt{2}} \dfrac{i}{\sqrt{2}} }^{2} +
        \abs{\dfrac{i}{\sqrt{2}} \dfrac{1}{\sqrt{2}} +\dfrac{1}{\sqrt{2}}
          \dfrac{i}{\sqrt{2}}}^{2} \\
       &= \dfrac{1}{4} + 1 = \dfrac{5}{4} = 1.25
\end{split}
\label{eqn:Measure_E_ideal}
\end{equation} 
\noindent
In this ideal scenario, therefore, the quantum measure associated with the event
$E$ takes a value exceeding unity, the maximum permissible value classically.

\paragraph{Experimental Setup:} 

The setup actually used in our experiment is diagrammed in Fig. \ref{fig:QMT_setup}. 
It differs from the idealized experiment just analyzed in two main respects.  
First,
we have replaced the Mach-Zehnder geometry (where the photon encounters two
identical beamsplitters) with a displaced Sagnac geometry (a DSI interferometer),
where the photon encounters the same beamsplitter twice;
and
second, we have used laser light rather than individual photons. \\

The change from MZI to DSI was done 
in order to mitigate noise from external vibrations.  
Such vibrations can affect the two paths
within the MZI differently, and this path-difference would cause the relative
phase between the two interfering beams (here, $01$ and $11$) to change over
time, thereby degrading the interference that enters into the intensity recorded
at the final detector.\\

 The choice to use laser light rather than a single-photon source
was also made to mitigate a sort of noise, in this case the ``statistical
noise'' that would arise from sending fewer photons through the apparatus.

Though the concept of event-filter, strictly speaking, applies to a single
photon, whose detection at the output port would count as a ``click'' in the
event-filter (thereby signalling the occurrence of the event), this does not
prevent an experiment using laser light from confirming that the value of the
quantum measure $\mu(E)$ has been obtained correctly.  
It's well known that (unless nonlinear optical elements are involved) the
intensity of laser light at an output port can serve as a valid proxy for the
probability that the photon would arrive there in a single photon experiment,
and this is all we need to conclude that our value of $1.172,$ is (within
errorbars) the same as what would be obtained from a stream of single photons.
We have provided an elementary derivation of this equivalence in the appendix.  
(For an analysis employing an over complete basis of coherent states, see \cite{Skagerstam2018} and \cite{Optical_Equivalence}.)\\

We turn now to the experimental details. \\

Light at wavelength $\lambda = 810 ~nm$ 
emitted from a narrow bandwidth
(linewidth typically $< 300 ~kHz$)
diode laser [$Toptica~ DL~ Pro$] 
provides
the input for this experiment. 
The laser beam is coupled to 
a $FC/PC$ to $FC/APC$ 
Polarization Maintaining Single Mode Fiber 
($PMSMF$) [$P5-780PM-FC-2, ~Thorlabs$] 
in order
$(i)$ to  maintain the polarization while
the beam propagates through the fiber, thus reducing the power
fluctuation after any polarization optics, 
$(ii)$ to reduce the 
pointing fluctuation 
about the transverse plane of the beam as compared to the
bare beam, 
$(iii)$ to render the spatial mode at the output fiber's output as
Gaussian as possible. 
The beam is collimated using an adjustable fiber collimator [$CFC11P-B, ~Thorlabs$]
that confines the divergence of the beam to less than $1~mrad$, 
and a mirror then redirects the beam toward $DSI$. 
%
In order to achieve a high degree of polarization purity,
a Glan Thompson Polarizer\footnote
{A Glan Thompson polarizer transmits the
 $s$-polarized component of beam (the $e$-$ray$) and reflects the
 $p$-polarized component of beam (the $o$-$ray$) of any unpolarized beam
 incident on it.} 
($GT$) [$GTH5M-B, ~Thorlabs$] 
with optic axis oriented to transmit 
the horizontal component of polarization ($\ket{H}$), 
is placed in the path of the beam 
between
the Collimator 
and
the $DSI$. \\


The horizontally polarized beam 
emerging from
the $GT$ is 
incident on the beam splitter 
$BS_{1}$ [$20BC17MB.2$, $~Newport$] 
which forms the Displaced Sagnac Interferometer ($DSI$) 
together 
with the mirrors $M_{T}$, $M_{R}$ and $M_{M}$ 
(all of them are [$5122, ~Newport$]). 
Let, the paths associated with 
the anticlockwise and clockwise directions of propagation 
of the beam inside the interferometer 
be
named as path-$A$ and path-$C$ respectively.
The state at $BS_{1}$ 
just after the 
first
pass 
is then
$\ket{\Psi_{1}} = \dfrac{1}{\sqrt{2}} (\ket{A}^{(0)} + i\ket{C}^{(1)})\ket{H}$%
\footnote{\interlinepenalty=10000 The superscripts on the spatial modes of the
  system give the path information, upper and lower paths being labelled as $0$
  and $1$ respectively}. 
The relative phase between the two paths of the
interferometer is controlled by tilting a Glass Plate $GP$ [$WG40530, ~Thorlabs$] 
in path-$A$. 
In one of the paths of the interferometer (here in path-$C$) 
a half-wave plate ($HWP_{1}$) [$WPO02-H-810-UM, ~Newlight Photonics$] 
is placed with its fast axis oriented at $45^{\circ}$ with respect
to the horizontal, 
thus realizing
the $\sigma_{x}$ evolution operator. 
Thereby,
the state at $BS_{1}$ just before the second pass becomes, 
$\ket{\Psi_{2}} = \dfrac{1}{\sqrt{2}} (e^{i\varphi_{g}}\ket{A}^{(0)}\ket{H} + i\ket{C}^{(1)}\ket{V})$. 
This implies that a
polarization measurement in the basis $\{\ket{H}, \ket{V}\}$ 
at one of the ports of the $DSI$ 
can give 
information about the path of the photon inside the interferometer. 
The state at $BS_{1}$ after second pass becomes 
$\ket{\Psi_{3}} = \dfrac{1}{2} (e^{i\varphi_{g}}\ket{U}^{(00)}\ket{H} + i^{2}\ket{U}^{(10)}\ket{V} + ie^{i\varphi_{g}}\ket{L}^{(01)}\ket{H} + i\ket{L}^{(11)}\ket{V})$, 
where
$\ket{U}$ and $\ket{L}$ respectively represent the paths corresponding to 
the upper and lower output ports of the $DSI$.  \\

After the $DSI$, actions on the polarization d.o.f. are performed to 
make the two histories $01$ and $11$ interfere, 
select the paths $00,01$, and $11$,
and
recombine them to ensure 
that
any detection at the position of \redlined{the power meter ($PM$)} as shown in Fig. \ref{fig:QMT_setup} 
corresponds 
to
the occurrence of the event $E$. A
polarizing beamsplitter $PBS_{1}$ [$PBS122, ~Thorlabs$] placed in path-$U$
reflects the history $10$ away from the setup,
 and a half-wave plate $HWP_{2}$
[$RZQ 2.15L.0810, ~B. ~Halle$] and $PBS$ ($PBS_{2}$) [$PBS122, ~Thorlabs$]
combination placed in path-$L$ makes the two histories $01, 11$ interfere which
otherwise 
would have had
orthogonal polarizations. 
$HWP_{2}$ has its fast axis oriented at
$22.5^{\circ}$ with respect to the horizontal; 
it
physically realizes a
Hadamard operator that transforms 
the $\{\ket{H}, \ket{V}\}$ basis 
to the $\{\ket{+}, \ket{-}\}$ basis 
\footnote{Here, $\ket{+} =
 \dfrac{1}{\sqrt{2}}(\ket{H}+\ket{V})$ and $\ket{-} =
 \dfrac{1}{\sqrt{2}}(\ket{H}-\ket{V})$}. 
Another half-wave plate $HWP_{3}$ [$RZQ 2.15L.0810, ~B. ~Halle$] 
acts
as
a
$\sigma_{x}$ operator, 
and 
a $50:50$ beamsplitter $BS_{2}$ [$BS005, ~Thorlabs$] 
placed in the upper arm ensures the beam from the
$00$ path ends up at $PM$ 
without any associated bias. \\ 

\paragraph{Data Acquision and Analysis:}
The $DSI$ is aligned in the collinear configuration, with a back-aligned
$GP$ being present in one of the paths (here, path-$A$). Next, the event
filter is setup by placing 
the various
polarization optics at 
their respective
locations,
 keeping in mind that only the beams from $00, 01, 11$ paths
should
be detected at the output 
and the beams from $01$ and $11$
need 
to be made to interfere. 
Using a power meter sensor $PM$ [sensor:
  $S121C, ~Thorlabs$, meter: $PM100D, ~Thorlabs$] (or a detector) the
powers are recorded 
($i$) at the input to the setup i.e., after the $GT$ (denoted $P_{I}$) 
and 
($ii$) at the output of the event filter (denoted $P_{E}$). 
Constructive interference is ensured by tuning the tilt
of the $GP$ in small steps while observing the power at the output and
fixing it at an angle that corresponds to maximum power. Other than
$P_{I}$ and $P_{E}$, the powers $P_{int}$, $P_{01}$, $P_{11}$, $P_{00}$
are monitored at the output of the event filter with (path-$U$),
(path-$U$, path-$C$) and (path-$U$, path-$A$), (path-$L$) being blocked
respectively, in order to get an idea about the interferometric phase
fluctuation and individual power fluctuations over time. \\

Experimental data is always associated with 
some 
limitations,
imperfections, noise, fluctuations, losses etc. 
In optical experiments
there are always 
losses associated with absorption in the material of the optical elements. 
Imperfections in the optical components being used, 
like deviations in the splitting ratios ($T:R$) of the beamsplitters from the quoted values, 
polarization dependent reflections\footnote%
{In
 most of the optical elements $R_{s} \ne R_{p}$, i.e., the reflectivity
 is not the same for $s-$ and $p-$ polarizations and adds a relative
 phase between $s-$ and $p-$polarized components making the beam
 elliptically polarized.}
from mirrors and beamsplitters, 
the extinction
ratios of the polarization optics, 
the surface quality of the components,
etc., 
modify 
the outcome of the experiment. 
Further, there are
systematic instrumental errors, 
like the nonlinear behavior of the power meter sensor, 
which are a matter of characterization. 
The 
laboratory 
conditions,
i.e., temperature, pressure, humidity, air current, etc. 
impact the optical alignment and the experimental data, 
causing 
a deviation in the experimental result from the theoretical value computed considering
ideal conditions. 
Additionally, in an interferometric experiment, the
phase instability inside the interferometer, wavefront distortion of the
beam, beam wander of the two overlapping beams, etc. 
will also lead to experimental errors. 
Hence, the experimental data needs to be analyzed
accounting for the potential sources of 
non-idealness
while providing
feasible corrections for some parameters 
depending on the degree they affect the experiment. \\  

\noindent\textbf{Experimental distribution of the quantum measure of $E$.}
In our implementation, the operational quantity associated with the event $E$ is the
probability that the event-filter registers a photon at its designated output port (PM).
With laser light (and in the absence of nonlinearities), this probability can be inferred from
the ratio of optical powers at the output and input of the filter. Accordingly, in an ideal
noise-free situation one would write
\begin{equation}
\mathcal{P}_{\rm exp}(E) \;=\; \frac{P_E}{P_I},
\end{equation}
where $P_I$ denotes the power incident on the setup (measured after the GT), and $P_E$ denotes
the power recorded at the output port of the event-filter.

In practice, neither $P_I$ nor $P_E$ is perfectly constant in time. Both exhibit (i) fast,
apparently random fluctuations (including high-frequency oscillations that can be reduced by
time-averaging), and (ii) slower systematic drifts (for example thermal or alignment drifts)
that are not removed by averaging over any single acquisition window. Since $P_I$ and $P_E$
are recorded at different times, simply combining non-simultaneous readings can bias the
estimate of $\mathcal{P}_{\rm exp}(E)$, and therefore of the inferred quantum measure.

To propagate these fluctuations into an uncertainty on $\mathcal{P}_{\rm exp}(E)$, we construct an
\emph{empirical distribution} of probabilities using Monte-Carlo resampling of the recorded
time series:
\begin{enumerate}
\item We record long time traces of the input and output powers, $P_I(t)$ and $P_E(t)$.
\item From each trace we draw (with replacement) a short-time dataset of duration
$\Delta t \approx 100~\mathrm{s}$, denoted $p_I$ and $p_E$ respectively. This window is long enough
to average over fast fluctuations, while still sampling the slow drifts across the full dataset.
\item For each draw we compute a probability estimate as the mean of the pointwise ratio,
\begin{equation}
\mathcal{P}^{(k)}_{\rm exp}(E) \;=\; \left\langle \frac{p_E}{p_I} \right\rangle ,
\end{equation}
and we repeat the procedure $N=10^{5}$ times to obtain the distribution
$\{\mathcal{P}^{(k)}_{\rm exp}(E)\}_{k=1}^{N}$.
\end{enumerate}

\smallskip
\noindent\emph{Correction for absorption and surface losses of $BS_1$.}
In the displaced Sagnac geometry the photon encounters the same $2$ inch beam splitter $BS_1$
twice. Absorption in the substrate (and additional surface losses) reduce the circulating power
and therefore reduce the measured $P_E$. We quantify this effect via the measured net power
throughput
\begin{equation}
\eta_s \;=\; \frac{P_T + P_R}{P_I} \;=\; T_{\rm abs} + R_{\rm abs},
\end{equation}
where $P_I$ is the power incident on $BS_1$, and $P_T$ and $P_R$ are the transmitted and
reflected powers from $BS_1$, respectively. Since the system passes through $BS_1$ twice, we
correct each resampled probability by
\begin{equation}
\mathcal{P}^{(c)}_{\rm exp}(E) \;=\; \frac{\mathcal{P}_{\rm exp}(E)}{\eta_s^2}.
\end{equation}

\smallskip
\noindent\emph{Filter success probability (postselection) and conversion to quantum measure.}
Even for an ideal, lossless implementation of the filter, the final stage has two complementary outcomes: only one corresponds to the monitored detector event at PM, while the other is routed to a complementary port that is not directed to the detector. In our implementation, the combination
$HWP_2$--$PBS_2$ (used to erase the polarization tag and enable interference of the $01$ and $11$
histories) together with the compensating beam splitter $BS_2$ in the $00$ arm routes the other
half to a complementary port that is not directed to the detector. Thus the factor of $1/2$ is a postselection success probability for the monitored outcome, not optical attenuation by $PBS_2$ (and it does not accumulate by inserting additional polarization beam splitters).
Equivalently, the event-filter’s detection probability is $\mu(E)/2$, so that the quantum measure
is obtained by multiplying by the corresponding factor of $2$:
\begin{equation}
\mu_{\rm exp}(E) \;=\; 2\,\mathcal{P}^{(c)}_{\rm exp}(E).
\end{equation}
Applying this mapping to the full distribution of $\mathcal{P}^{(c)}_{\rm exp}(E)$ yields the
experimental distribution of $\mu_{\rm exp}(E)$ reported in Fig.~5.

\paragraph{Experimental corrections to the theoretically expected value of the measure of $E$:}
The theoretical formula given in Eqn. \ref{eqn:Measure_E_ideal} computes
the value of the quantum measure for the event $E = \{00,01,11\}$ of a
photonic system to be $\mu_{th}^{ideal}=1.25$, considering the ideal
system, devices, and laboratory conditions. Any loss in any part of the
setup would reduce the amplitudes associated with different paths which
would effectively lower the value of quantum measure obtained
experimentally. Also, the computation of $\mu_{th}^{ideal}$ considers
constructive interference between the paths $01$ and $11$, i.e., the
relative phase between the two paths of the interferometer to be
$\varphi = 0$. Hence, any variation in phase from zero would only reduce
the obtained interference intensity and effectively reduce the value of
$\mu(E)$. Therefore, it is important to have an estimation of the range
within which the experimentally obtained quantity would lie, given the
non-ideal laboratory setting. 

For event $E=\{00,01,11\}$, the modified theoretical expression for
measure considering the effect of different parameters associated with
the experiment is given as, 
\begin{equation}
\begin{split}
\mu^{e}_{\varphi}(E) &= \abs{A_{e}(00)}^{2} \\
                    &\quad + \abs{A_{e}(01)+\exp(i\varphi) A_{e}(11)}^{2}
\end{split}
\label{eqn:mu_E_modified}
\end{equation}
\noindent
The second term in the above expression represents the interference
intensity of $01,11$ beams in presence of interferometric phase
variation. The amplitudes $A_{e}(\gamma)$ for different paths $\gamma$
are computed considering the transmission of the system through the
optical components present in the respective paths in the setup. This
includes taking into account the overall transmission factor ($\eta$)
through the components, splitting ratio $T:R$ ratio of the
beamsplitters, the extinction ratio of the polarization optics, possible
changes in the polarization due to the $HWP$ misalignment\footnote{The
  uncertainty associated with the $HWP$ orientation, here, is related to
  the random detection noises and random power variations as the fast
  axis of a $HWP$ is not aligned looking at the label of the rotation
  mounts but from the observation of powers at the output ports of a
  $PBS$ placed after it.}, 
and polarization dependent reflectivity
$R_{p} \ne R_{s}$ of the mirrors used etc. The variation in the
relative phase $\varphi$ is determined as, 
\begin{align}
	\varphi = \arccos{\left(\dfrac{I(\varphi) - I_{1} - I_{2}}{2 \sqrt{I_{1} I_{2}}}\right)}
	\label{eqn:phase_unequal}
\end{align}
\noindent
provided the following condition is satisfied,
\begin{align}
	I(\varphi) \le I_{1}+I_{2}+2 \sqrt{I_{1}I_{2}}
	\label{eqn:criteria_interference}
\end{align}
\noindent
Here, the intensities $I(\varphi)$, $I_{1}$ and $I_{2}$ are respectively
determined from the recorded power data $P_{int}$, $P_{01}$ and
$P_{11}$. A distribution for the phase $\Phi$ is obtained by choosing
samples randomly from $I(\varphi)$, $I_{1}$ and $I_{2}$, and determining
$\varphi$ for which the criteria in Eqn. \ref{eqn:criteria_interference}
is satisfied. Since, determination of the phase depends on the
distribution of the powers, $\Phi$ does not only represent the phase
variation, but possible effect due to power fluctuation as well. The
distribution 
$\Phi$ and distribution of the amplitudes 
$A_{e}(00), A_{e}(01), A_{e}(11)$ associated with the inherent uncertainty in the
parameters of the real components, gives a range of possible
$\mu^{e}_{\varphi}(E)$ values resulting in a distribution $\mu_{th}(E)$,
for the event $E$.\\ 

\section{\label{sec:Results}Results and Statistical \redlined{Analysis}}

\begin{figure*}[t]
	\centering
	\includegraphics[width=0.99\linewidth]{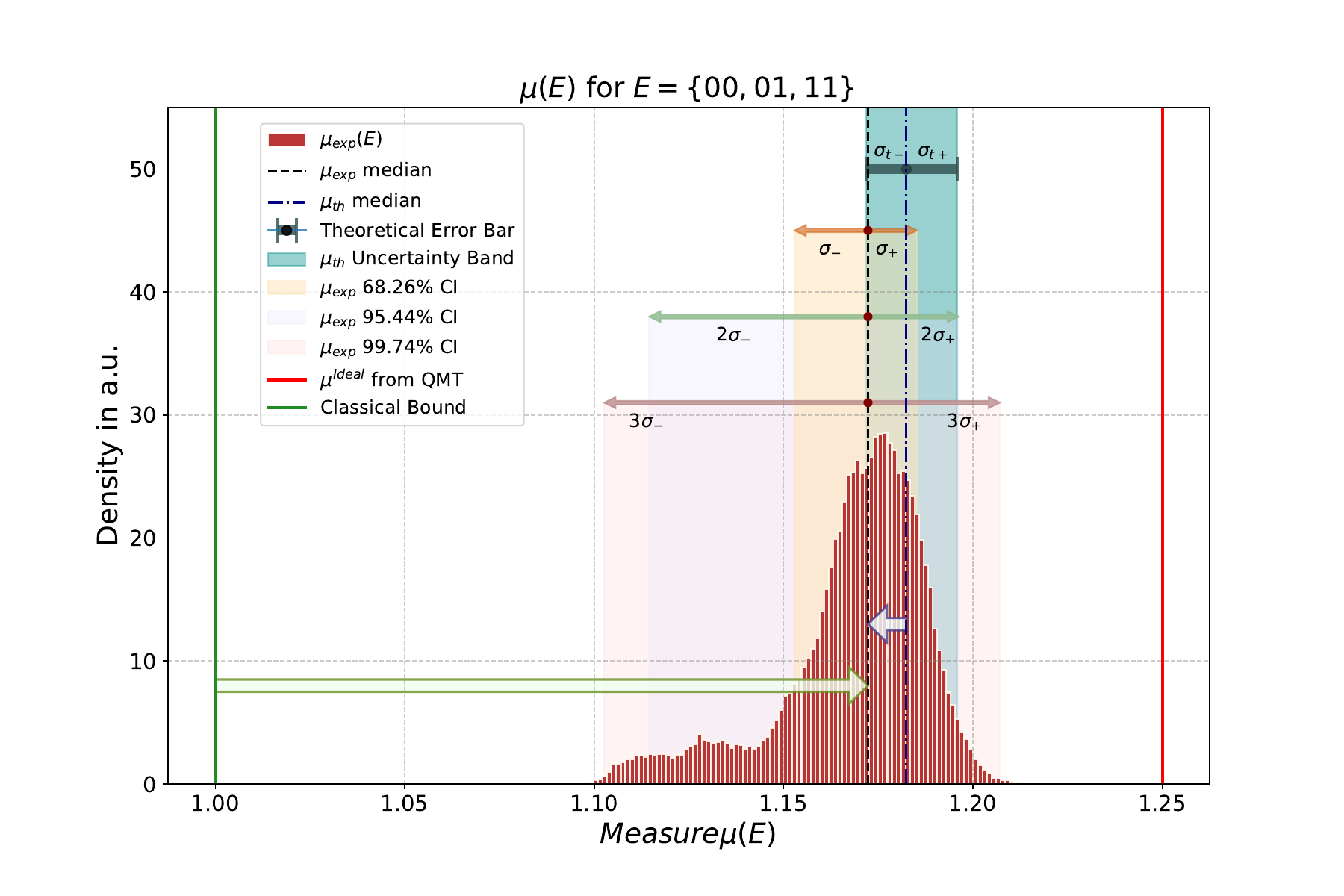}
	\caption{Experimental result for the \emph{experimental} non-serial event $E_{\rm exp}=\{00,01,11\}$ implemented with the event-filter of Fig.~4 (Section 4). Shown is the histogram of the empirically inferred distribution of the quantum measure $\mu_{\rm exp}(E_{\rm exp})$ obtained from the measured input/output powers of the event-filter (Section 5). The shaded regions indicate the central $68.26\%$, $95.44\%$, and $99.74\%$ confidence intervals of the (generally non-Gaussian) distribution. The dashed vertical line marks the experimental median. The blue vertical line and blue band indicate the median $\mu_t$ and $1\sigma$ uncertainty of the theoretical distribution $\mu_{\rm th}(E_{\rm exp})$ obtained after propagating experimental imperfections (Section 5); the black marker shows the corresponding theoretical error bar. The red line denotes the ideal theoretical value $\mu^{(\mathrm{ideal})}(E_{\rm exp})=5/4$, and the green line marks the classical bound $\mu_{C(\max)}=1$. (The pedagogical example of Section 3 uses a different illustrative event and is not the one shown here.)}

	\label{fig:mu_final}
\end{figure*}

The histogram plot for the experimentally obtained distribution of
measure $\mu_{exp}(E)$ for the event $E=\{00,01,11\}$ of the photonic
system is shown in Fig.\ref{fig:mu_final}. The distribution is
asymmetric in nature, with the asymmetry mostly arising from the
systematic temporal drifts of the power data $P_{E}$ and $P_{I}$,
recorded at different times, that are used in the determination of
$\mu_{exp}$. \redlined{Because $\mu_{\rm exp}(E)$ is asymmetric, we summarize its central value by the median and its spread by the lower and upper percentile widths defined below.} \\

\paragraph{\redlined{Percentile-based separations for an asymmetric distribution.}}
Since $\mu_{\rm exp}(E)$ is generally non-Normal and asymmetric, we quantify
(i) the central value by the median $\mu_e$ of the sampled distribution, and
(ii) the spread by the percentile-based offsets $\sigma_{\pm}$, where
$\mu_{84.13}$ and $\mu_{15.87}$ denote the $84.13^{\mathrm{th}}$ and $15.87^{\mathrm{th}}$
percentiles of the sampled distribution, respectively:
\begin{align}
\sigma_{+}=\mu_{84.13}-\mu_e,\qquad \sigma_{-}=\mu_e-\mu_{15.87}.
\label{eq:sigma_pm}
\end{align}
{\color{black}
We compare the experimental median $\mu_e$ with a reference value $\mu_0$, chosen as either the classical bound $1$ or the median $\mu_t$ of the theoretical prediction including experimental imperfections. Using the fixed experimental percentile widths $\sigma_+$ and $\sigma_-$ defined in Eq.~\eqref{eq:sigma_pm}, we report both normalized separations:
\begin{equation}
\begin{aligned}
S_+(\mu_0)&=\frac{\mu_e-\mu_0}{\sigma_+},\\
S_-(\mu_0)&=\frac{\mu_e-\mu_0}{\sigma_-}.
\end{aligned}
\label{eq:percentile_separations}
\end{equation}
The subscripts $+$ and $-$ specify the upper and lower widths about the experimental median, not a classification of samples relative to $\mu_0$; both ratios describe the same signed displacement using different widths. They are descriptive quantities, not Gaussian $z$-scores or tail-probability estimates.\par}

For the
event $E=\{00,01,11\}$, we report the experimentally obtained value of
measure 
to be $\mu_{e}(E): 1.172^{+0.013}_{-0.019}$. 
The $68.26\%$,
$95.44\%$ and $99.74\%$ confidence intervals ($CI$) associated with the
$1\sigma$, $2\sigma$ and $3\sigma$ \redlined{percentile-based} regions of the distribution, are presented
with three different shaded regions in the plot. The theoretical
expectation of measure under ideal 
laboratory
conditions is shown with a red
line at $\mu^{(ideal)}(E) = 1.25$. Considering imperfections, losses,
power fluctuation, phase variation and other uncertainties associated
with different parameters that can impact the experimental outcome, the
value of the expected measure is obtained to be
$1.182^{+0.013}_{-0.011}$, representing the median and $\pm 1\sigma$
error of the distribution $\mu_{th}(E)$. The vertical blue line and the
blue band in the plot respectively represent the $median$ and the
$1\sigma$ uncertainty of $\mu_{th}(E)$. This blue band can be considered
as theoretical uncertainty band. The median (central) theoretical value
of measure corresponding to $\mu_{th}(E)$ is denoted by $\mu_{t}$.

\bigskip
Considering only the effects of the real optical components present in
the paths and disregarding the power and phase fluctuations while
assuming $\varphi = 0$, the theoretically expected value of measure for
the event $E=\{00,01,11\}$ is obtained to be $1.2095 \pm 0.0015$. Next,
considering the effects of real components and the power variation while
ignoring phase fluctuation, the expected theoretical value of measure is
computed to be $1.1981^{+0.0084}_{-0.0075}$ from the expression below, 
\begin{equation}
\begin{split}
\mu^{e}_{\varphi=0}(E) &= I_{e}(00) + I_{e}(01) + I_{e}(11) \\
                     &\quad + 2 \sqrt{I_{e}(01)~I_{e}(11)}
\end{split}
\label{eqn:mu_E_formula_Intensity}
\end{equation}
\noindent
Here, the intensities $I_{e}(\gamma)$ are obtained from the experimentally recorded data, $P_{00}, P_{01}, P_{11}$ from the following expression, where $P_{\gamma}$ denotes the experimentally recorded power corresponding to the history $\gamma \in \{00,01,11\}$ (i.e., $P_{\gamma} \in \{P_{00},P_{01},P_{11}\}$ in this work),
\begin{align}
	I_{e}(\gamma) = \dfrac{2}{\eta_{s}^{2}}\left(\dfrac{P_{\gamma}}{P_{I}}\right)
	\label{eqn:Intensity_end}
\end{align}

{\color{black}
Relative to the classical probability bound $\mu_{C(\max)}=1$, the experimental median has normalized separations $S_+(1)=13.32$ and $S_-(1)=8.89$, expressing the same excess above unity in units of the upper and lower experimental percentile widths respectively, implying that the quantity $\mu_{e}$ is non-classical in nature. Indeed, for comparison with the theoretical median $\mu_t$ computed within the framework of Quantum Measure Theory (QMT) for the actual laboratory conditions, the absolute separations are $|S_+(\mu_t)|=0.78$ and $|S_-(\mu_t)|=0.52$, and the theoretical median lies within the experimental central $68.26\%$ percentile interval.\par}
Unlike a classical measure, a quantum measure 
incorporates quantum interference and therefore can take values greater than one. \\


\section{\label{sec:Discussion}Discussion}

In an experiment that begins with a ``preparation'' and ends with a
``registration'', these two events are ``initial'' and ``final'' only to
the extent that we agree to ignore events which took place either
earlier than the first or later than the second; this we ordinarily do
for the sake of simplicity. But why should we ignore as well those
events that happened at ``in-between'' times? Even if, for the sake of
argument, we grant that such \textit{in-between} events are inaccessible
to observation, they nevertheless acquire a meaning in the context of
history-based frameworks such as quantum measure theory. \\

Consider for example the particular event $E$ which is the subject of
our experiment, and whose verbal description could be 
``the photon did not follow path $10$''. 
The known initial and final locations of the
photon do not by themselves determine whether or not $E$ happened, but
that does not mean that nothing can be said on the matter. In a
stochastic world, one would not expect to be able to deduce with
certainty whether $E$ happened, but what theory does provide in the
classical case is a \textit{probability} for $E$ to have happened. In
the quantum case, it provides a kind of ``non-classical probability'' or generalized measure for $E$: its \textit{quantum measure}, which in the present instance is
$\mu(E)=5/4$.\\

But how exactly is one to understand the physical meaning of this
number? On being presented for the first time with the concept of
\textit{quantum measure}, someone might well ask whether $\mu(E)$ has
any direct experimental significance outside of the special case where
$E$ is an instrument-event and $\mu(E)$ can therefore be interpreted as
a classical probability. \\
 
If, as we hope, theoretical developments based on the quantum measure
can help to resolve the interpretational paradoxes of quantum mechanics
(if $\mu$ can find a role in a more complete description of the
micro-world), this would amply answer the question being raised. We are
not there yet, but a big step in that direction would be any way to put
$\mu(E)$ on a more immediate experimental footing. A procedure to do
precisely that is what reference \cite{Frauca_2017} provided. Indeed, it
implicitly showed, 
via the concept of what we are calling an \textit{event filter}, 
how to bring not only the \textit{measure} of an event, 
but also in a certain sense the \textit{event concept} itself, 
into direct contact with experiment. 
In effect, reference \cite{Frauca_2017} put forward two main claims. \\

The first claim was that $\mu(E)$ can indeed be measured experimentally.
For that purpose, it introduced an ancilla-based protocol that
could in principle furnish an event-filter for an arbitrary event $E$.  
However, 
the photonic event $E_{\rm exp}=\{00,01,11\}$ of our experiment 
makes a particularly interesting test case.
Its measure exceeds unity, 
and therefore admits of no interpretation as a probability in the classical sense.  
Moreover $E$ appears to correspond to no selfadjoint operator, 
and it does not seem possible to determine $\mu(E)$ by any
sequence of simple measurements of projective or POVM type.
Thus, none of the more familiar measurement procedures seem applicable
to this event.
However, the protocol of \cite{Frauca_2017} does apply.\\

Inspired by this protocol, 
our experiment employs a design, 
which however differs from that of \cite{Frauca_2017} in that 
the ``ancillas'' are in some sense integrated into the ``system'', with
the photon's polarization contributing the necessary extra degrees of freedom.
In relation to \cite{Frauca_2017} the resulting event-filter improves
the ``counting efficiency'' from 1/3 to 1/2.  We think that one can
trace the reason for this improvement to the fact that 
in place of two separate ancillas that couple only once to the system, 
our design uses in effect a single ancilla that couples twice to the system.\\

Within experimental error, our event-filter correctly determined the measure of event $E$ to be 5/4,
and in particular to be
greater than $1$ \redlined{by $13.32$ upper or $8.89$ lower experimental percentile widths}. Our table-top experiment uses only commonly employed
  pieces of optical apparatus, and it would not be difficult to
  predict its outcome by the ordinary procedure of evolving a
  state-vector via the Schr{\"o}dinger equation. However, such a
  computation would fail utterly to bring out how the apparatus acts
  as a filter for a very simple and definite ``{in-between}'' event $E$
  --- a specific set of three photon trajectories. A comment on ``just interference'' is in order. Of course the optical network can be modeled and its output intensities predicted within standard quantum optics. What is nontrivial is that, for a genuinely non-serial \emph{system event} $E$, no single output-port intensity of the unfiltered system corresponds to $\mu(E)$, because the contributing histories do not define a single-time observable. The role of the event-filter is precisely to implement an operational yes/no question about $E$ by converting it into a detector event $D$ with ordinary probability $P(D)\le 1$, and with a calibration factor relating $P(D)$ to $\mu(E)$.\\

Although our experiment employed full-power laser light as the source,
its results would apply equally to an ensemble of single photons,
because an interference pattern generated from a coherent source of
light (exhibiting wave nature) is equivalent 
to the average pattern
produced by an ensemble of single photons (exhibiting corpuscular nature).
This equivalence is discussed further in the appendix, 
where it is seen to rest on very elementary features of quantum field theory. \\

The second main claim implicit in reference \cite{Frauca_2017} was that
the event-filters proposed therein are, in a well-defined sense,
\textit{non-destructive} (exactly as the name ``filter'' suggests). By
this we mean that when the detector ``clicks'', the resulting
``collapsed wave function'' of the ``system'' (in this case the photon) is
whatever results when one evolves the initial wave function by summing
over precisely those paths that correspond to the event $E$. (In
particular, the interference terms between pairs of these paths are not
disturbed.)\footnote{In this way, one obtains a technique for
  state-preparation that might conceivably be useful in applications
  like quantum computing.} \\

Because our experiment ends up absorbing the photon, 
it could not address this second claim directly.

However a pair of relatively simple modifications would allow it to do so.

Firstly, in order to avoid destroying the photon, one could (in the spirit of a
negative-result measurement) remove the detector from the main output port and
instead install detectors on all the remaining output ports.  
A positive outcome or ``click'' would then be signalled by these complementary
detectors \textit{not registering} the photon, which therefore would survive to
be subjected to further tests.
Secondly, one would need to replace the laser with a single-photon source so
that the distinction between ``click'' and ``no click'' would become
meaningful. 
Both of these modifications to the experimental setup would be straightforward,
but they would make the collection of adequate statistics more tedious, and they
would require a re-consideration of the loss factor depending on the detection
efficiency of the single-photon detectors. \\

Then, in order to test whether the event filter was functioning as
claimed, one would need to introduce, downstream from the filter per
se, some type of ``state tomograph'' designed to verify that the
effective state of the emergent photon was the one determined by the
propagation-amplitudes exhibited in equation \ref{eqn:mu_E_formula}.\\

Imagine then that these modifications have been made and the experiment
carried out. What will we have learned thereby about the micro-world? At
an intuitive level, one would like to believe that when ``the detector
clicks'', it is informing us that event $E$ has actually happened.
Appealing as it is, this conclusion must remain in abeyance until the
notorious ``quantum foundational'' puzzles are more fully resolved than
they are at present. However, if it is ultimately upheld it will answer,
in the particular case of our experiment, the question this paper began
with: What are we measuring when we perform a quantum measurement?.

\section{\label{sec:Conclusion}Conclusion}
This paper presents an experiment that demonstrates how
the \textit{quantum measure} of a photonic event can be determined, not
only theoretically but practically.  It does this for the specific experimental event,
$E_{\rm exp}=\{00,01,11\}$,
and confirms with near certainty that its \redlined{experimentally inferred} measure $\mu(E_{\rm exp})$ exceeds
the maximum value permissible for a classical event (namely probability = 1).\\

The experiment is performed by devising an optical setup which
selects for the photon paths that belong to $E_{\rm exp}$.  Equivalently it
``filters out'' the photon paths that do not belong to $E_{\rm exp}$, which it
does by arranging for the photon's polarization to tag them for
rejection.
In this way, it provides a non-destructive procedure allowing inferences
about the intermediate processes that a micro-system undergoes during
its evolution from preparation to detection.\\

\redlined{For $\mu(E_{\rm exp})$, the experiment reports $1.172^{+0.013}_{-0.019}$. The theoretical median, calculated accounting for the laboratory conditions and imperfections in the optical elements, lies within the experimental central $68.26\%$ percentile interval. The experimental median exceeds the classical probability bound of $1$ by $13.32$ upper or $8.89$ lower percentile widths.}
It thus
bears witness to a large amount of constructive
interference between the histories (photonic trajectories) that comprise
the event~$E_{\rm exp}$.\\

By bringing the framework of Quantum Measure Theory into the laboratory,
our experiment enhances QMT's empirical support, hopefully paving the
way for foundational studies that could give us a physical model of
microscopic events.
Such a model would answer the pivotal question posed earlier: What
happened in between the initial emission and the final detection of the
photon?\\

The quantum measure can be thought of as a sort of generalized Born rule
that attributes generalized probabilities (possibly exceeding unity) to
non-instrument events that are experimentally inaccessible by
conventional quantum measurement.
In relation to such an event, an event-filter of the type reported here
can be thought of as a sort of \textit{active Fresnel zone plate} that 
selects
a family 
of light-rays which a purely passive zone-plate cannot access.
Beyond its purely theoretical implications, such a filter potentially
opens up opportunities for innovative quantum circuits and new protocols
in the field of quantum computing and quantum information more
generally.

\Needspace{13\baselineskip}
\section{Author Contributions}

S.C. performed the experiment, acquired the data, and carried out the
analysis. R.D.S. formulated the theory underlying the experiment, building
on his earlier work on quantum measure theory. U.S. conceived, designed,
and supervised the experiment and worked closely with R.D.S. on developing
its theoretical aspects. All authors contributed to writing the manuscript.

\section{Acknowledgements}

U.S. is grateful for the hospitality of Perimeter Institute where part of
this work was carried out. 
Research at Perimeter Institute is supported
in part by the Government of Canada through the department of
Innovation, Science and Economic Development Canada and by the Province
of Ontario through the Ministry of Economic Development, Job Creation
and Trade. This research was also supported in part by the Simons
Foundation through the Simons Foundation Emmy Noether Fellowship Program
at Perimeter Institute. 
U.S. also acknowledges partial support provided
by the Ministry of Electronics and Information Technology (MeitY),
Government of India under a grant for Centre for Excellence in Quantum
Technologies with Ref. No. 4(7)/2020-ITEA as well as partial support from the National Quantum Mission of the DST.

\section*{\label{sec:appendix}APPENDIX -- Relating multi-photon to single-photon experiments}

\def\expect#1{\langle #1 \rangle}
\def\alfa{\alpha}

It is generally accepted that in an experiment like ours, the intensity of
laser light at a particular output port can serve as a valid proxy for
the probability that the photon would arrive there in a single photon
experiment.
%
%
In our case, this means
%
%
that the ratio $P_E/P_I$ can be interpreted as the probability that a single
photon released by the laser would arrive at the photodetector PM.
Since this relationship was presupposed by our discussion 
in the main text,
we derive it here.  In fact we provide two
derivations.  The first derivation models the source as a coherent state; the
second derivation generalizes the first to a much broader class of sources.\\

Given that our derivations are elementary, using little more than the
first properties of creation and annihilation operators, we hope they
will hold some independent interest.
For an analysis employing an overcomplete basis of coherent states, see \cite{Skagerstam2018}. \\

It seems simplest to couch our analysis in terms of wave-packets or
``modes'', which propagate from source to detector.  To each mode there
will correspond an annihilation operator $a$ and a creation operator
$a^*$, and also a \textit{coherent state} $\ket{\alpha}$, which
generalizes to field theory the description of a minimum-uncertainty
wave-packet oscillating back and forth in a harmonic potential.  Such a
state provides (when highly excited) an optimal approximant to a
solution of the classical equations of motion of the electromagnetic field
(or of any other bosonic field that one might be considering).
The two modes of interest will be that emitted by the source and the
``sub-mode'' thereof which arrives at the desired output after passage
through the apparatus.  We will designate them as `$a$' (input) and
`$b$' (output) respectively, and normalize the associated
creation/annhilation operators in the usual manner suited to a particle
representation, $[a,a^*]=[b,b^*]=1$.\\

Consider first the case of a one-particle state such as a single photon
source would emit, namely $a^*\ket{0}$ (where $\ket{0}$ denotes the vacuum). 
The amplitude that this photon would be received at the output port is
given by the inner product, 
$\bra{0}ab^*\ket{0}$,
of 
$a^*\ket{0}$ with 
$b^*\ket{0}$, 
the state that represents a single photon arriving at the port.  
The probability of reception
is thus the absolute square of $\bra{0} a b^* \ket{0}$.  
Because $a$ annihilates the vacuum, 
we can also write 
$\bra{0} a b^* \ket{0}$ 
as
$\bra{0} [a,b^*] \ket{0}$ or simply as $[a,b^*]$, 
since $\ket{0}$ was normalized to $\expect{0|0}=1$.
Here we have used the fact that
in a linear (``free'') field theory,
 a commutator like $[a,b^*]$ reduces to a $c$-number.
The probability we're after is therefore 
\begin{equation} p_E = |[a,b^*]|^2 \end{equation}

Consider next the case where the source emits a multiparticle coherent state 
\begin{equation}
            \ket{\psi} = \ket{\alfa} = \exp\{\alfa a^*\} \ket{0}
\end{equation}
(where for simplicity we work with the un-normalized state-vector).
How many photons would be present, on average, at the input and output ports?
The answer is given by the expectation-values of the respective number operators
$a^*a$ and $b^*b$, which are 
\begin{align}
\label{eqn:PIE}
  P_I &= \bra{\psi} a^* a \ket{\psi} / \expect{\psi | \psi}  \quad \hbox{and} \\ \nonumber
  P_E &= \bra{\psi} b^* b \ket{\psi} / \expect{\psi | \psi} \ .
\end{align}

We can compute both at once if we notice that 
for \textit{any} pair of annihilation operators,
$a$ and $b$, 
and any parameter $\alfa$,
one has
\[
\begin{aligned}
b \exp\{\alfa a^*\}\ket{0}
   &= [b, \exp\{\alfa a^*\}] \ket{0} \\
   &= [b, {\alfa a^*}] \exp\{\alfa a^*\}\ket{0} \\
   &= \alfa [b, a^*] \exp\{\alfa a^*\}\ket{0}
\end{aligned}
\]
where the second equality used the fact 
that $[b,\cdot]$ is a \textit{derivation},
and also that, as a $c$-number, 
$[b,{\alfa a^*}]$ commutes with the exponent $\alfa a^*$.
Substituting $b=a$ 
on one hand and $b=b$ 
on the other hand 
furnishes the pair of equations 
\begin{equation} a\ket{\alfa} = \alfa [a, a^*] \ket{\alfa} = \alfa\ket{\alfa} \end{equation}
and
\begin{equation} b\ket{\alfa} = \alfa [b,a^*] \ket{\alfa}\ \end{equation}
The first of these equations is nothing but the defining equation for a coherent state.
It says that $\ket{\alfa}$ is an eigenvector of $a$ with eigenvalue $\alfa$.
The second equation says that $\ket{\alfa}$ is \textit{also} an eigenvector of $b$,
but with the eigenvalue $[b,a^*]\alfa$.  
(Thus, a coherent state is an eigenvector of \textit{every} annihilation operator.) \\

Now it's true in general that 
$\bra{\psi} a^* a \ket{\psi} = \expect{a\psi | a\psi}$,
the squared norm of $a\ket{\psi}$.
But since for $\ket{\psi}=\ket{\alfa}$ both $a\ket{\psi}$ and $b\ket{\psi}$ are multiples of $\ket{\psi}$,
it's immediate that the squared ratio of their norms is the squared ratio of these multiples.
In other words,
\begin{equation}
  P_E/P_I = |[b,a^*] \alfa|^2 / |\alfa|^2 = |[b,a^*]|^2
\end{equation}
precisely the single-photon probability $p_E$ we found above!
(Of course $p_E$ could also have been written as $p_E/p_I$, $p_I$
denoting the probability\footnote%
  {In these equations, we have used small $p$ to suggest ``probability'' and big $P$ to suggest ``power''.}
to find the photon at its source, which is unity by definition.
For completeness, we may also mention that the separate values for $P_I$ and $P_E$
in equations (\ref{eqn:PIE}) 
would have been
$P_I = |\alfa|^2 / \expect{\alfa|\alfa}$ and 
$P_E = |\alfa|^2 |[b,a^*]|^2 / \expect{\alfa|\alfa}$.)\\

This completes our demonstration, but one might wonder, Why did it
work?, and one might ask, What about other input states beside just a
single photon or a coherent state of them?  As it happens, the answer to
the first question leads to a theorem that answers the second question
for a much larger class of states $\psi$, including states with a definite
number of photons. \\


\noindent{\bf Proof of a more general equivalence}

We preserve the notations from above.
In particular $\psi$ (which we now take to be normalized to $\expect{\psi | \psi} = 1$) will be the state ``emitted by'' our source.
The expected number of particles (photons) received at the output port $b$ is then
\begin{equation}n_b = \expect{\psi| b^* b |\psi} = ||b \psi||^2 \ ,\end{equation}
while the average number emitted by the source is similarly
\begin{equation}n_a = \expect{\psi| a^* a |\psi} = ||a \psi||^2\end{equation}

A review of the earlier demonstration brings to light the feature that the
state-vectors which entered into it always came in pairs, $\psi$ cum
$\psi'$, so related that the application of any annihilation operator to
$\psi$ produced a scalar multiple of $\psi'$, thereby reducing the whole
calculation to a comparison of these multiples.  And if one looks
further for the reason behind this feature, one can identify it as being
the fact that the vectors $\psi$ always had 
the form, $\psi = F(a^*)\ket{0}$. 
We claim that for any vector $\psi$ of this form the ratio 
$P_b/P_a$ assumes the same value, 
namely the absolute square of $[b, a^*]/[a, a^*]$.\\

The proof of this claim will basically repeat our earlier derivation, 
where now we set
\begin{equation}
    \psi = \ket{F} =  F(a^*)\ket{0}
\end{equation}
for some 
(sufficiently regular)
function $F(x)$, 
normalized such that 
$\expect{F|F}=1$.
Applying the annihilation operators $a$ and $b$ to $\ket{F}$, we find
\begin{align}
   a \ket{F} &= a F(a^*) \ket{0} = [a, F(a^*)] \ket{0} \nonumber\\
              &= [a, a^*] F'(a^*)\ket{0} \\
   b \ket{F} &= b F(a^*) \ket{0} = [b, F(a^*)] \ket{0} \nonumber\\
              &= [b, a^*] F'(a^*)\ket{0}
\end{align}
where the prime on $F$  denotes differentiation.
Consequently 
$\psi = \ket{F}$
does indeed have a
companion-vector $\psi'$, which we may take 
to be
\begin{equation} \psi' = F'(a^*)\ket{0}\ , \end{equation} 
so that
\begin{equation} 
  a \, \psi = \lambda_a \, \psi'  \qquad  b \, \psi = \lambda_b \, \psi'
\end{equation}
with
\begin{equation}  \lambda_a = [a, a^*]  \qquad  \lambda_b = [b, a^*]  \end{equation}
Then the ratio $n_b/n_a$ is simply
\begin{equation} 
   \frac{n_b}{n_a} = \frac{|\lambda_b|^2 ||\psi'||^2} {|\lambda_a|^2 ||\psi'||^2} = \left|\frac{\lambda_b}{\lambda_a} \right|^2 = \left|\frac{[b , a^*]} {[a , a^*]} \right|^2
\end{equation}
Therefore
\begin{equation} 
    n_b/n_a =  |[b , a^*] / [a , a^*]|^2 \ ,
\end{equation}
a ``universal'' ratio%
\footnote{Of course $\lambda_a [a, a^*]=1$ for us, because we have normalized $a$ that
  way, but we have retained $\lambda_a$ and $[a, a^*]$ because, (1) it is sometimes convenient to
  use a different normalization, and (2) writing things this way preserves the parallelism between $a$ and $b$. }
that is entirely independent of the choice of the free function  $F$.
This holds in particular: \\

$\bullet$ for $F(x)=x$ (single-photon experiment),

$\bullet$ for $F(x)\propto \exp(\alfa x)$ (laser-like coherent-state experiment), and 

$\bullet$ for $F(x)\propto x^N$ (experiment with precisely $N$ photons, all occupying the same mode).\\

Notice here that our proof has absorbed the single-photon case into the
case of general $F$.  
That doing so was legitimate is easily established
by noticing first that 
the choice, $F(x)=x$, yields $\psi = F(a^*)\ket{0} = a^*\ket{0}$,
and then observing further that 
when $\psi=a^*\ket{0}$, 
the expectation $\expect{\psi|n_b|\psi}$ 
is nothing but the probability for the photon to be
present at the output port (because%
\footnote{or more formulaically, because $\expect{\psi|n_b|\psi} = |\expect{\psi | b^*\ket{0}}|^2 $}
the random variable $n_b$ can then only assume the values 0, 1).\\

\noindent
Remark. The derivation would not change if we were to replace `$a$' in `$P_a$' by
some third annihilation operator (perhaps belonging to some other
fraction of the original beam, exiting from a different output port).
We would then find 
for the ratio of powers at two distinct output ports:
\begin{equation} 
    n_b/n_c = |[b , a^*] / [c , a^*]|^2 \ , 
\end{equation} 
which also follows from the equality
$\lambda_b/\lambda_c=(\lambda_b/\lambda_a)(\lambda_c/\lambda_a)$.\\

To conclude this appendix, let us record here a lemma, abstracted from
the above proofs, that isolates the formal mathematical basis of our
derivations.  We will state it in terms of ``annihilation operators''
and their adjoints, the associated ``creation operators''.
By annihilation operators we will mean a collection of operators which:
annihilate a vector $\ket{0}$ called the vacuum; 
commute mutually among themselves; 
and 
yield a complex multiple of the identity operator when commuted with any creation operator.\\

\noindent
LEMMA. Let $A$ be a set of annihilation operators, and let $a$ be an
element of $A$. Let $F(z)$ be a sufficiently regular%
\footnote{We have stated the lemma in the same informal manner as we
 have adopted for the whole of this appendix, but we have included the
 phrase ``sufficiently regular'' in order to indicate that some
 condition on $F$ of differentiability or holomorphicity would be needed in any
 rigorously formulated version of the lemma.}
complex function of
$z$, and set $\ket{\psi}=F(a^*)\ket{0}$.  Then for any $b,c\in A$
\begin{equation}
   \frac {\expect{\psi| b^* b |\psi}} {\expect{\psi| c^* c |\psi}}           
   = 
   \left| \frac {[b,a^*]} {[c,a^*]} \right|^2
\end{equation}

\bibliographystyle{quantum_measure_doi}
\bibliography{Quantum_Measure_2026-09-14_Quantum_refs}

\end{document}